\documentclass{aa}

\usepackage{graphicx}
\usepackage{placeins}
\usepackage{txfonts}
\usepackage{natbib}
\usepackage{subfigure}
\begin{document}

   \title{Bars \& boxy/peanut bulges in thin \& thick discs}

   \subtitle{I. Morphology and line-of-sight velocities of a fiducial model}
   \author{F. Fragkoudi
          \inst{1}
          \and
          P. Di Matteo\inst{1}
           \and
          M. Haywood\inst{1} 
          \and
          A. G\'{o}mez\inst{1} 
          \and
          F. Combes\inst{2,3} 
          \and
          D. Katz\inst{1} 
          \and
          B. Semelin\inst{2} 
          }

   \institute{GEPI, Observatoire de Paris, PSL Research University, CNRS, Univ Paris Diderot, Sorbonne Paris Cit\'{e}, Place Jules Janssen, 92195,\\
Meudon, France\\
              \email{francesca.fragkoudi@obspm.fr}
         \and
             Observatoire de Paris, LERMA, CNRS, PSL Univ., UPMC, Sorbonne Univ., F-75014, Paris, France
         \and
             College de France, 11 Place Marcelin Berthelot, 75005, Paris, France
             }

   \date{}

 
  \abstract
   {We explore trends in the morphology and line-of-sight (los) velocity of stellar populations in the inner regions of disc galaxies, using N-body simulations with both a thin (kinematically cold) and a thick (kinematically hot) disc which form a bar and boxy/peanut (b/p) bulge. The bar in the thin disc component is $\sim$50\% stronger than the thick disc bar and is more elongated, with an axis ratio almost half that of the thick disc bar. The thin disc b/p bulge has a pronounced X-shape, while the thick disc b/p is weaker with a rather boxy shape. This leads to the signature of the b/p bulge in the thick disc to be weaker and further away from the plane than in the thin disc. 
Regarding the kinematics, we find that the los velocity of thick disc stars in the outer parts of the b/p bulge can be \emph{larger} than that of thin disc stars, by up to 40\% and 20\% for side-on and Milky Way-like orientations of the bar respectively.
This is due to the different orbits followed by thin and thick disc stars in the bar-b/p region, which are affected by the fact that: i) thin disc stars are trapped more efficiently in the bar - b/p instability and thus lose more angular momentum than their thick disc counterparts and ii) thick disc stars have large radial excursions and therefore stars from large radii with high angular momenta can be found in the bar region.
We also find that the difference between the los velocities of the thin and thick disc in the b/p bulge ($\Delta v_{los}$) correlates with the initial difference between the radial velocity dispersions of the two discs ($\Delta \sigma$) . 
We therefore conclude that stars in the bar - b/p bulge will have considerably different morphologies and kinematics depending on the kinematic properties of the disc population they originate from.}


   \maketitle
%

\section{Introduction}

   Bars are ubiquitous features in the local universe, with about two thirds of disc galaxies containing bars with variable strengths \citep{Eskridgeetal2000,Menendezetal2007,Barazzaetal2008,Aguerrietal2009,Gadotti2009}.
Boxy/peanut/X-shaped (b/p) bulges (also referred to as boxy/peanuts, or b/p's) are structures which extend out of the plane of nearly half of all edge-on disc galaxies in the local universe \citep{Luttickeetal2000}. Numerical simulations and orbital structure analysis have shown the intimate link between bars and b/ps, by demonstrating that once a bar forms, a b/p will likely form soon after (e.g. \citealt{Combesetal1990,MartinezValpuestaetal2006}), since b/p's are caused by vertical orbital instabilities, which cause the bar to puff out vertically from the plane of the disc (e.g. \citealt{Binney1981,PfennigerFriedli1991,Skokosetal2002, Portailetal2015}). Recent work has also shown how b/p's can affect the evolution of disc galaxies, by reducing the bar-driven gas inflow (\citealt{Fragkoudietal2015, Fragkoudietal2016}; for reviews on these topics the reader is referred to \citealt{Athanassoula2016} and references therein.) 

Furthermore, from extensive photometric and spectroscopic studies of the Milky Way bulge, we now know that the inner region of the Milky Way contains a bar and a b/p \citep{Weilandetal1994, Dweketal1995, Howardetal2009, Natafetal2010,McWilliamandZoccali2010,NessLang2016}. Indeed, recent results suggest that the \emph{main} component of the Milky Way bulge is a b/p (Ness et al. 2012, 2013; \nocite{Nessetal2012,Nessetal2013a} \citealt{WeggGerhard2013}), with estimates placing an upper limit on the mass of a possible classical bulge between $\sim$2-10\% \citep{Shenetal2010, Kunderetal2012, DiMatteoetal2014,Kunderetal2016,Debattistaetal2016}.

Due to the ubiquitous nature of thick discs in observed \citep{Burstein1979, Tsikoudi1979,YoachimDalcanton2006} and simulated galaxies \citep{Abadietal2003,Birdetal2013,Stinsonetal2013,Martigetal2014}, along with the various formation scenarios proposed for them \citep{Quinnetal1993,Brooketal2004,SchoenrichBinney2009,Quetal2011,Martigetal2014,Minchevetal2015,Haywoodetal2015}, the question of the interplay between such a population and structures such as bars and b/p's naturally arises. 
In the Milky Way in particular, due to the small scalelength of the $\alpha$-enhanced thick disc population \citep{Bensbyetal2011,Bovyetal2012}, it follows that the chemically defined thick disc is centrally concentrated; this, along with recent results from chemical evolution models which indicate that the mass of the thick disc could be of the same order as that of the thin disc \citep{Haywoodetal2013, Snaithetal2015}, point to the fact that the thick disc will have an important contribution in terms of mass in the central region of the Milky Way. There have also been recent claims that (geometrically defined) thick discs make up a significant fraction of the baryonic content of galaxies \citep{Comeronetal2011}.
Therefore the question of how thick discs are mapped into a bar and a b/p bulge is of foremost interest to our understanding of the structure and evolution of disc galaxies in general, and to the Milky Way in particular.

In this study we therefore explore how thin and thick discs are mapped into a bar and a b/p bulge. This work is further motivated by the fact that, with the exception of few recent studies, not much work has been done on the composite nature of bars and b/p bulges in galaxies with both a thin and a thick disc, whether these are modelled as discrete components or with a continuum of populations (see e.g. \citealt{BekkiTsujimoto2011, BekkiTsujimoto2011b, DiMatteo2016} for models with discrete stellar populations and e.g. \citealt{Debattistaetal2016,Athanassoulaetal2017} for models with a continuum of stellar populations.) It is important to note that when employing models with discrete stellar populations, as in the case of the models presented here, we do not necessarily imply that galaxies are made up of two distinct components (for example, studies of stellar populations in the Milky Way have shown that it has a disc whose properties vary continuously with scaleheight -- see \citealt{Bovyetal2012}). Modelling galaxies with discrete components essentially helps to simplify the problem, while also allowing for controlled simulations (in which one can vary the scalelenghts, scaleheights and kinematic properties at will), and lends itself for understanding the trends that will be followed by stellar populations with different kinematic properties.

While we know that stars in the disc will get trapped in the bar instability as a function of how kinematically hot the population is (as was found by studies of individual discs e.g. \citealt{Hohl1971,AthanassoulaSellwood1986,Combesetal1990,Athanassoula2003}) there has not been a systematic study on how the properties of bars and b/p bulges will vary as a function of the properties of the kinematically hot/thick disc component, such as its mass and scalelength.  We will therefore address the dynamics, morphology and kinematics of bars and b/p bulges in simulations with both thin and thick discs, of various masses and scalelengths, in a series of papers (Fragkoudi et al. in prep.) in order to understand how bars map stellar populations with different dynamical properties into the inner regions of galaxies. In this paper -- the first in the series -- we explore a fiducial simulation, focusing on the main trends in morphology and line of sight velocities of stars in bars and b/p's in setups with both thin and thick discs. We focus on giving physical interpretations for these observed trends, based largely on the exchange of angular momentum the two populations are subject to.

The structure of the paper is as follows: in Section \ref{sec:5models} we describe the simulations and the initial conditions. In Section \ref{sec:morph} we discuss the basic morphological features of our fiducial model with both a thin and a thick disc, and we show how these discs are mapped into the bar and b/p bulge. In Section \ref{sec:results} we focus on the kinematics, specifically on the line of sight (los) velocity of the models, and show that the los of the thick disc can be \emph{higher} than that of the thin disc in the b/p region. In Section \ref{sec:mapping} we explore how the angular momentum redistribution and the radial velocity dispersion affect the morphology and the los velocities of the thin and thick disc stars in the bar - b/p. In Section \ref{sec:discussion} we discuss observational studies relevant to these results and in Section \ref{sec:summary} we summarise the main conclusions of the paper.

\section{The simulations \& Initial Conditions}
\label{sec:5models}
\begin{table*}
\centering
\label{tab:info}
\begin{tabular}{ r | c | c | c | c | c | c | c | c | c | c} 
Property & $r_D$ & $h_z^{thin}$  & $h_z^{thick}$ & $n_{disc}^{thin}$ & $n_{disc}^{thick}$ & $m_d$ & $r_H$ & $n_{halo}$ & $m_h$ & $\epsilon$ \\ \hline
Value & 4.7\,kpc & 0.3\,kpc & 0.9\,kpc & 700000 & 300000 & 9.2$\times$10$^4$$M_{\odot}$ & 10\,kpc & 500000 &  3.2$\times$10$^5$$M_{\odot}$ & 150\,pc   \\
\end{tabular}
\caption{Properties of the fiducial simulation used in this study. From left to right the columns correspond to the characteristic radius of both the thin and thick disc, the characteristic height of the thin disc, the characteristic height of the thick disc, the number of particles in the thin disc, the number of particles in the thick disc, the mass of the disc particles, the characteristic radius of the dark matter halo, the number of particles in the dark matter halo, the mass of the halo particles and the softening length.}
\end{table*}

In this study we use an N-body simulation which has both a thin and a thick disc, with characteristic scaleheights of 0.3 and 0.9\,kpc respectively to study disc galaxies with stellar populations with different kinematic properties. The model is of course a simplification of reality, since it is unlikely that most disc galaxies contain two very well-separated and distinct components, as is evidenced by studies of stellar populations in the Milky Way, which have shown that it has a disc whose properties vary continuously with scaleheight (e.g. \citealt{Bovyetal2012}). This discretisation reduces the complexity of the model and lends itself for understanding the trends that will be followed by the stellar populations in the Milky Way and external disc galaxies. 

\begin{figure}[]
\centering
\includegraphics[width=0.98\linewidth]{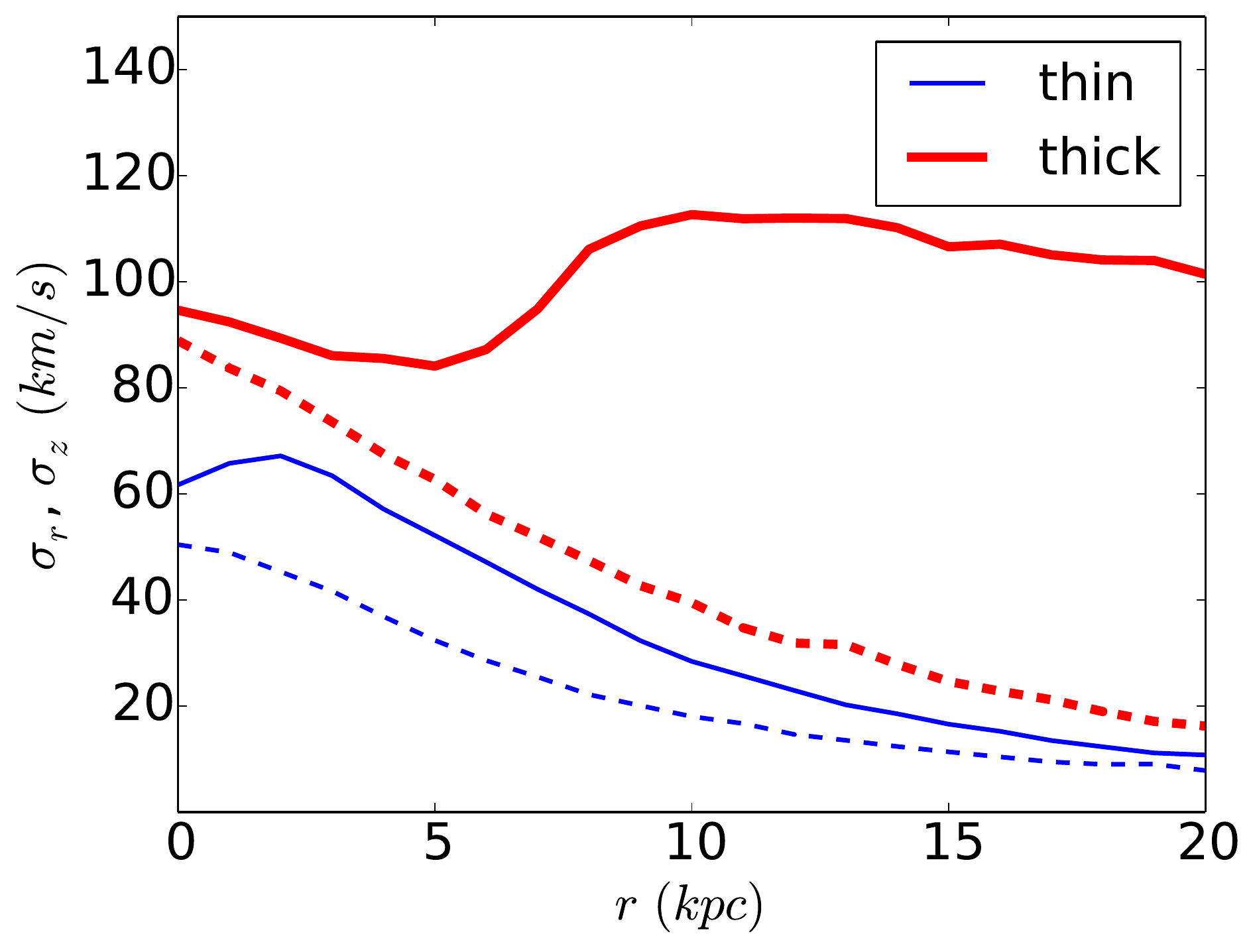}
\caption{Radial (solid) and vertical (dashed) velocity dispersion profiles of the thin (blue) and thick (red) discs in the initial conditions of the fiducial model used in this study.}
\label{fig:dispIC}
\end{figure}

The initial conditions of the discs explored in this study are obtained using the algorithm of \cite{Rodionovetal2009}, the so-called ``iterative'' method.
The algorithm constructs equilibrium phase models for stellar systems, using a constrained evolution so that the equilibrium solution has a number of desired parameters. As we do not have tight constraints of the radial and tangential velocity dispersion of galaxies in the local Universe (and even less so at higher redshifts) we do not impose a profile to these velocity dispersions, instead allowing the system to converge to an equilibrium solution. The only constraining parameter is the density distribution of the two discs, which are described by a Miyamoto-Nagai profile \citep{BT2008} with a characteristic radius $r_D$ of 4.7\,kpc; the velocity dispersion is let to evolve unconstrained, with the requirement that the initial conditions (ICs) generated are in equilibrium. At the end of the ``iterative'' method we obtain velocity dispersion profiles for the thin and thick discs as shown in Figure \ref{fig:dispIC}, which are correspondingly kinematically cold and hot, also in terms of the radial velocity dispersion. Note that the vertical velocity dispersion profile naturally converges to that expected from equation,
\begin{equation}
\sigma_z  (r)= \sqrt{c \pi G \Sigma(r) h_z}
\end{equation}

which is derived from the Poisson and Jeans equations, and which relates the vertical distribution of stars ($h_z$) and their vertical velocity dispersion ($\sigma_z$) to the mass distribution ($\Sigma$) as a function of radius $r$ (where the constant $c$ depends on the form of the height function -- see \citealt{vanderkruit1988,vanderKruitFreeman2011}).

The thick disc mass is 30\% the total baryonic mass (i.e. of both the thin and thick disc combined, where the baryonic mass is $M_{\star}$ = 1$\times$10$^{11}$$M_{\odot}$). The number of all the disc particles is $n_{bar}$ = 1$\times$10$^6$, each with mass $m_D$ = 9.2$\times$10$^4$ $M_{\odot}$. The dark matter halo is modelled as a Plummer sphere \citep{BT2008} with mass $M_H$ = 1.6$\times$10$^{11}$$M_{\odot}$, characteristic radius $r_H$ = 10\,kpc, and $n_{halo}$ = 5$\times$10$^5$ particles, each with mass 3.2$\times$10$^5$$M_{\odot}$. In Table \ref{tab:info} we summarise some of the main properties of the simulations used in this study.

To run the simulations we employ the Tree-SPH code of \cite{SemelinCombes2002}, in which gravitational forces are calculated using a hierarchical tree method \citep{BarnesHut1986}; for a full description of the code the reader is referred to \cite{SemelinCombes2002}. As there is no gas in these simulations the SPH part of the code is not employed and the gravitational forces are calculated using a tolerance parameter $\theta$=0.7 and include terms up to the quadrupole order in the multipole expansion. A Plummer potential is used to soften gravity at scales smaller than $\epsilon$ = 150\,pc. The equations of motion are integrated using a leapfrog algorithm with a fixed time step of $\Delta t$ = 0.25\,Myr.

In the snapshots taken from the end of the simulation, after 9\,Gyr of evolution, once a bar and a b/p forms -- which are those used for most of the analysis in the rest of this paper -- we renormalise the bar length to be 5\,kpc, similar to values found for the Milky Way and external galaxies \citep{Gadotti2011, Weggetal2015}. We measure the bar length using two methods: 1) by eye, where the bar length is taken as the distance between the centre of the galaxy to the outer tip of the bar component, usually where the spiral arms begin (e.g. \citealt{Martin1995}) and 2) using the ellipse fitting method, where the bar length is measured as the semi-major axis of maximum ellipticity in the bar (e.g. \citealt{Erwin2005}). The difference between these two measurements is of the order of 20\%, as found in previous studies (e.g. \citealt{Martin1995,Erwin2005}) and we adopt the average value between the two methods.

\section{Morphology of thin and thick disc bars and b/p's} 
\label{sec:morph}

\begin{figure*}
\centering
\includegraphics[width=0.9\textwidth]{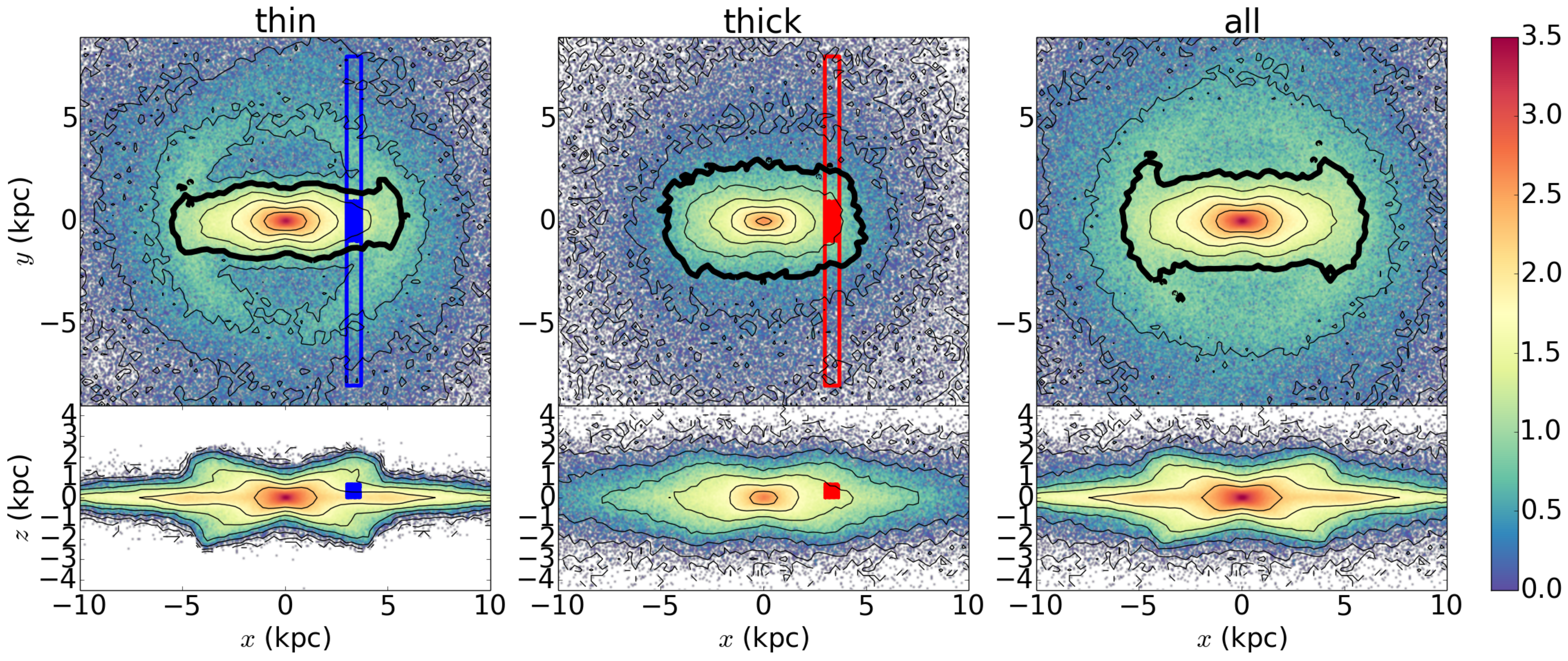}
\caption{Surface density maps for the thin and thick disc stars (first and second columns respectively) and all the stars in the disc (third column). The top panels show the $xy$ (face-on) projection and the bottom panels show the $xz$ (edge-on) projection of surface density. The blue and red shaded and unshaded regions indicate the selected particles at the end of the bar discussed in Section \ref{sec:mapping} and in Figures \ref{fig:lz} and \ref{fig:birth} (see text).}
\label{fig:xysurf}
\end{figure*}

\begin{figure}
\centering
\subfigure[bar strength]{%
	\includegraphics[width=0.46\linewidth]{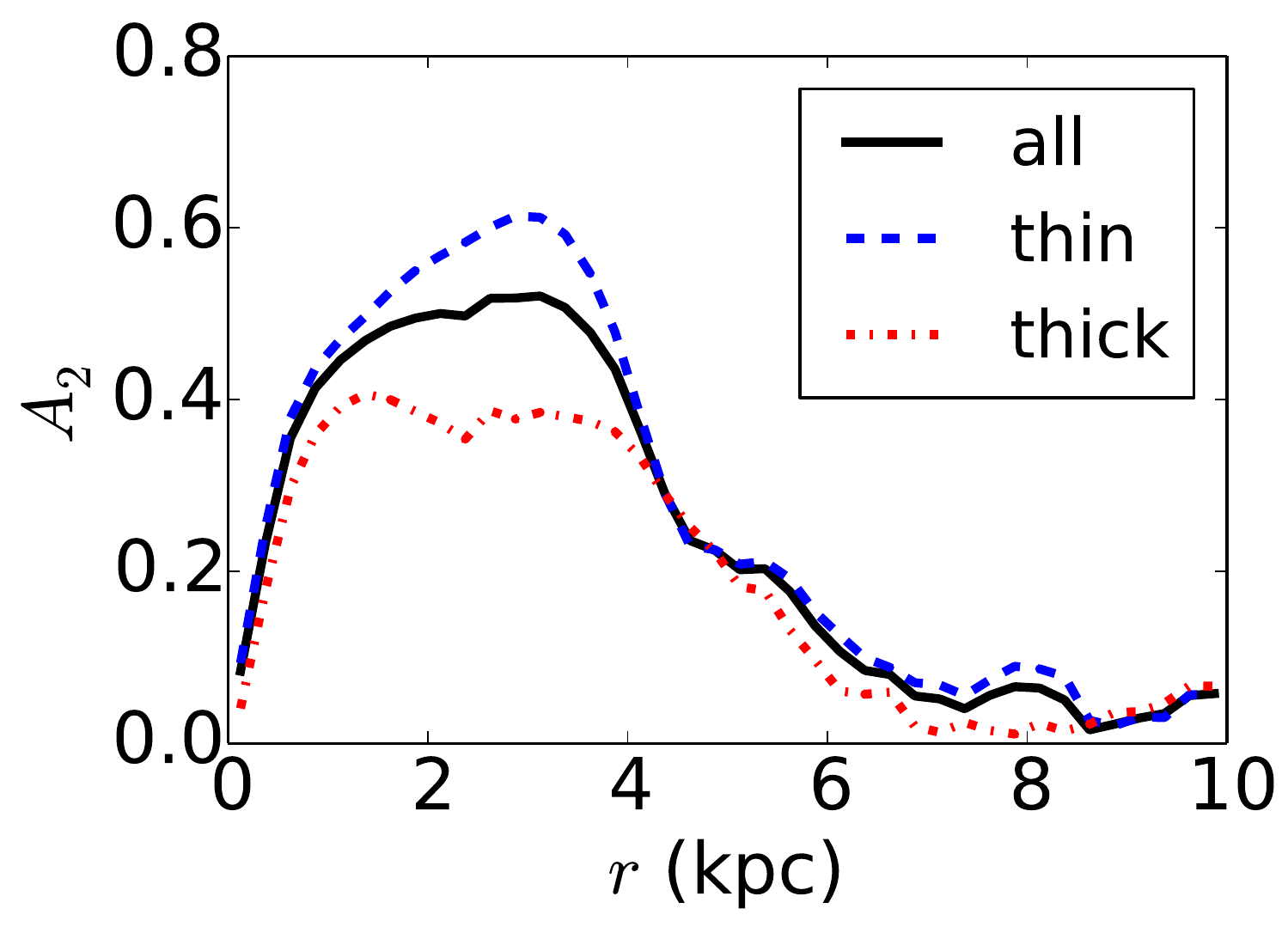}
	\label{fig:bars}}
\quad
\subfigure[b/p strength]{%
	\includegraphics[width=0.46\linewidth]{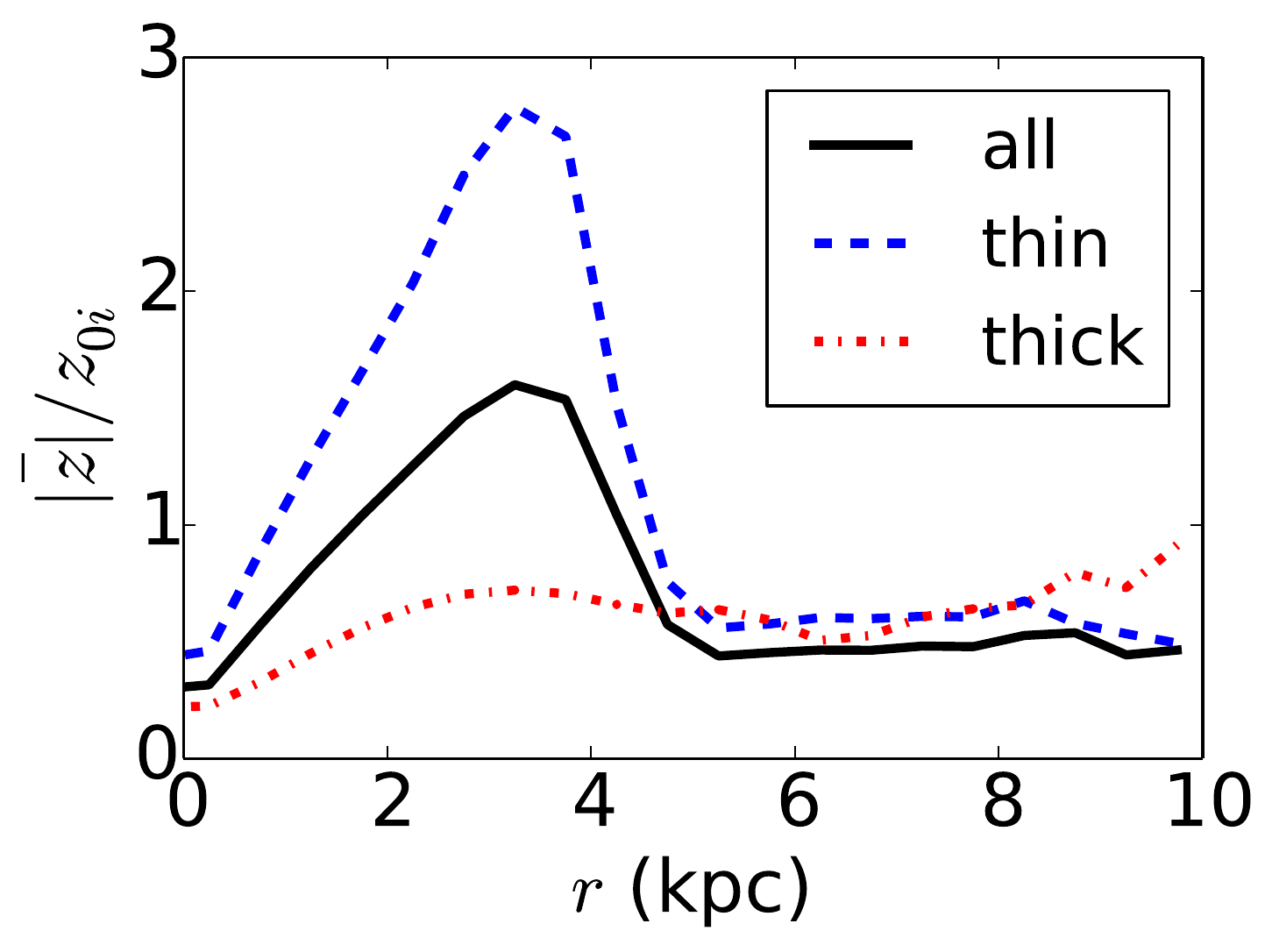}
	\label{fig:pea}}
\quad
\caption{(a) The bar strength $A_2$ as a function of radius at the end of the simulation, for the thin (dashed blue line) and thick (dashed-dotted red line) discs separately and the total bar strength (solid black line). We see that the thick disc bar is weaker than the thin disc bar. (b) The b/p strength (see Equation \ref{eq:bps}) of the thin and thick disc b/p and the total b/p bulge strength (colours as in plot (a)).}
\label{fig:morph}
\end{figure}

\begin{figure*}
\centering
\subfigure[$z$ = 0.2\,kpc]{%
	\includegraphics[width=0.31\linewidth]{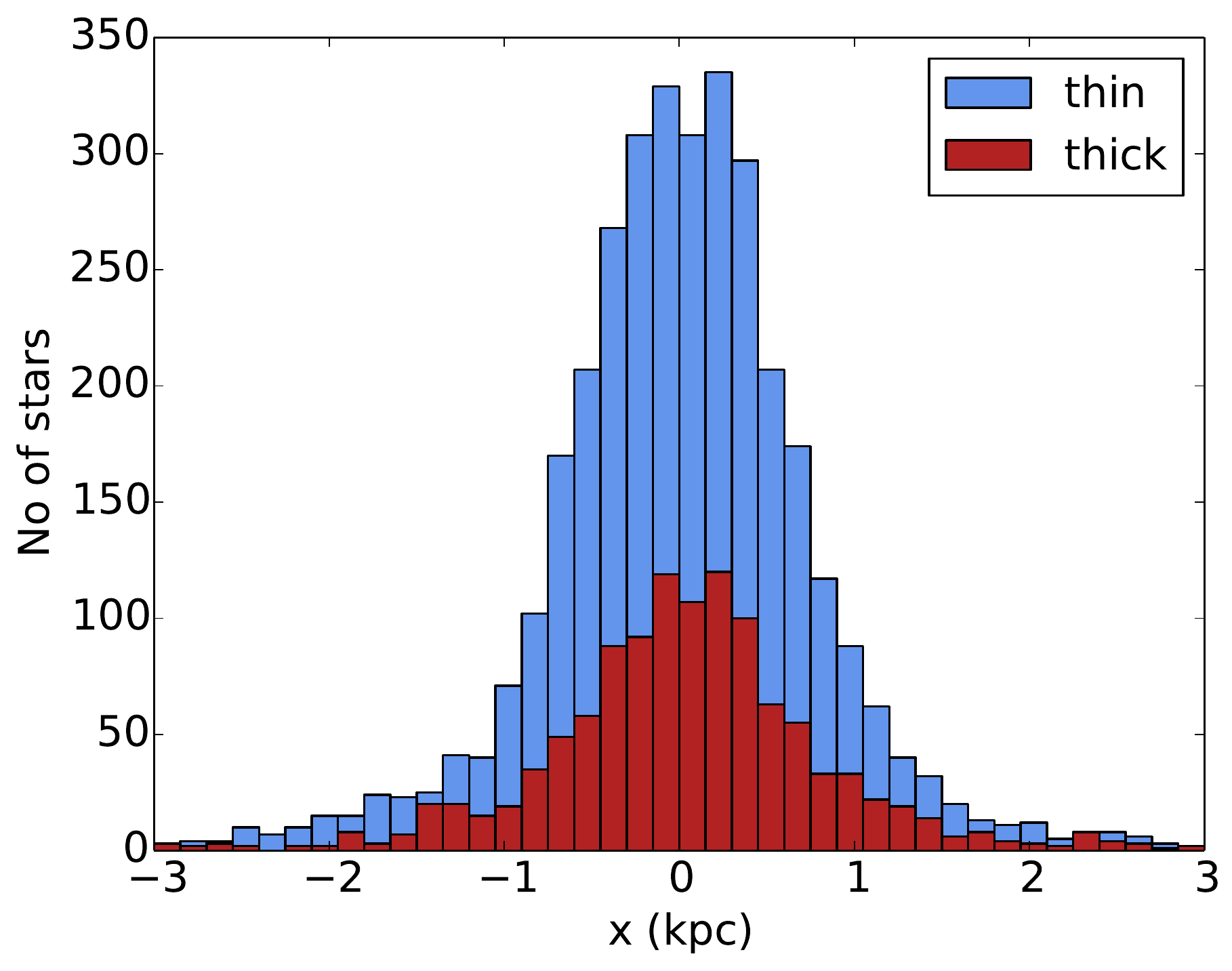}
	\label{fig:los_dens_lowlow}}
\quad
\subfigure[$z$ = 0.4\,kpc]{%
	\includegraphics[width=0.31\linewidth]{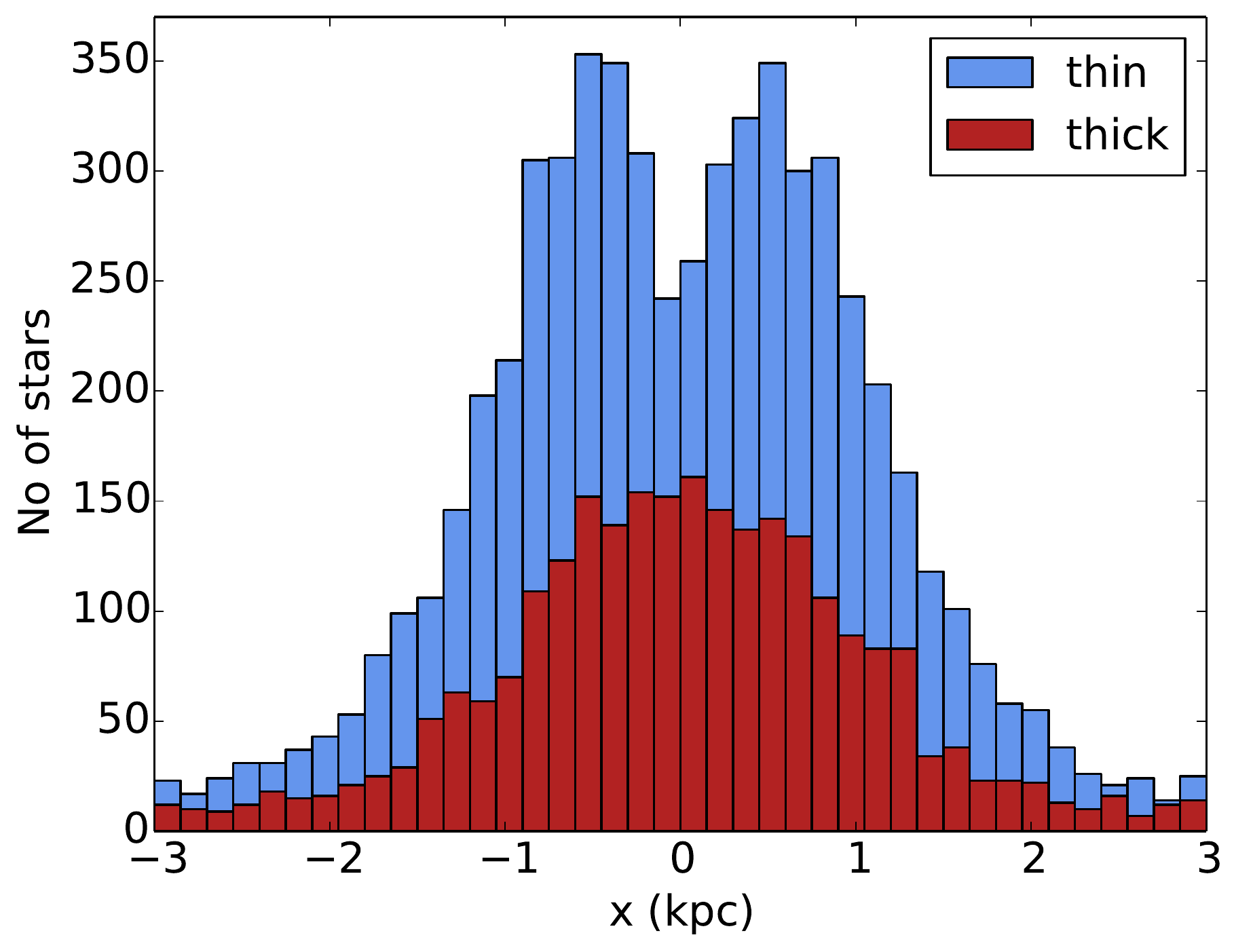}
	\label{fig:los_dens_low}}
\quad
\subfigure[$z$ = 0.6\,kpc]{%
	\includegraphics[width=0.31\linewidth]{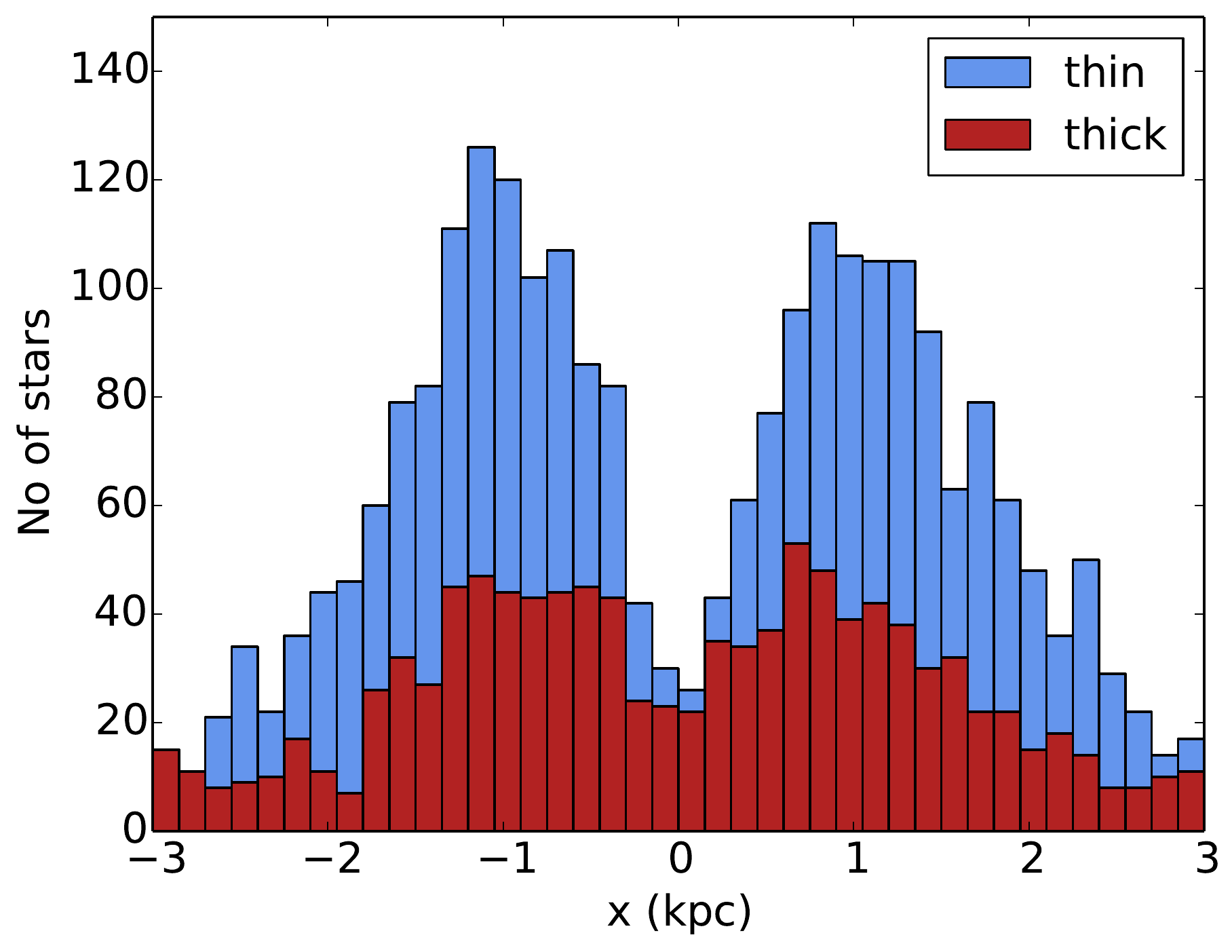}
	\label{fig:los_dens_high}}
\quad

\caption{The distribution of stars along the bar major axis at different heights above the plane, starting from close to the plane on the left and moving further away from the plane towards the right. We see that the signature of the b/p bulge is weaker and appears at larger heights above the plane for the kinematically hotter component.}
\label{fig:losdens}
\end{figure*}

\begin{figure}
\centering
\includegraphics[width=0.9\linewidth]{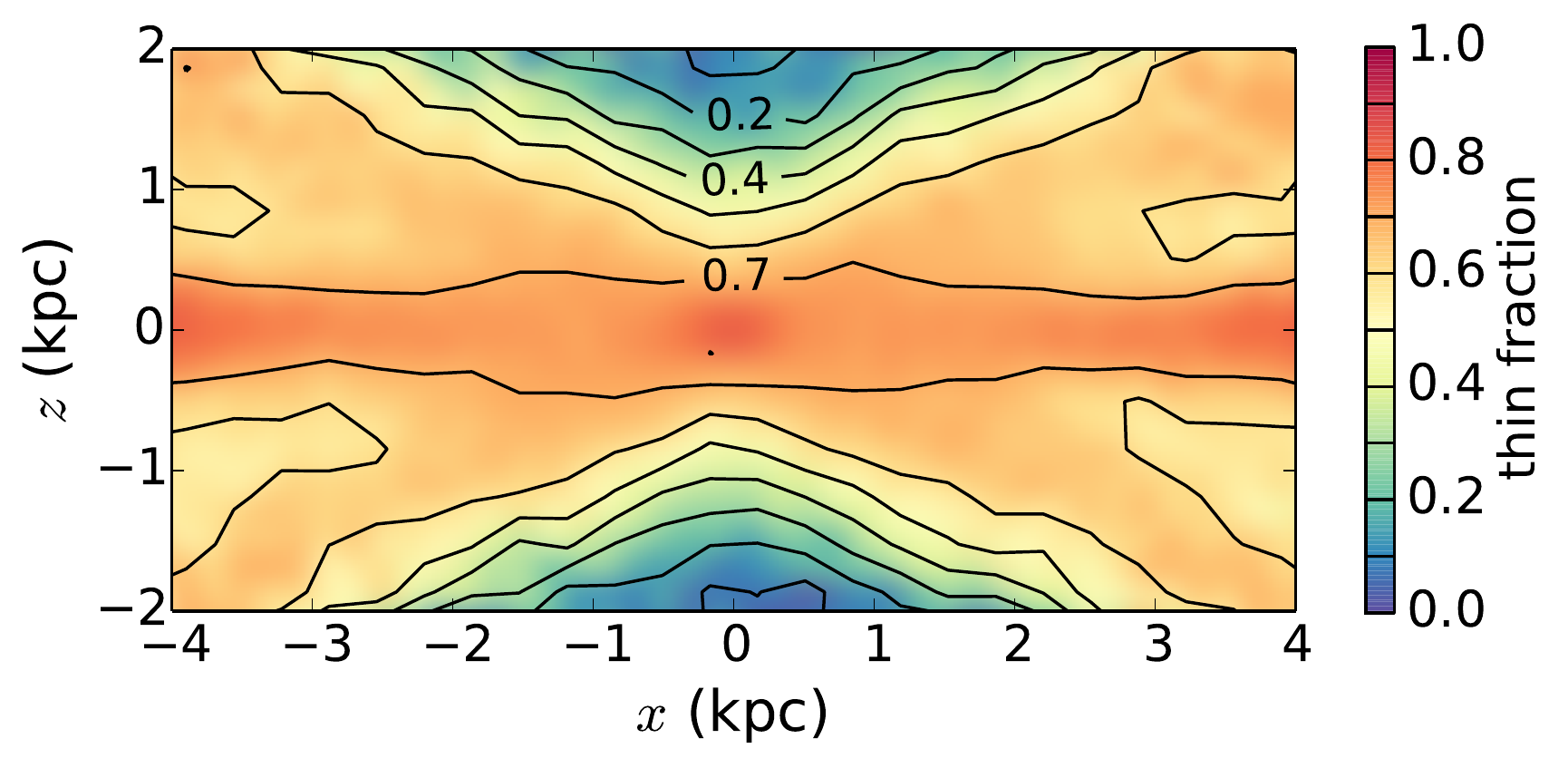}
\caption{The fraction of thin disc stars in the $xz$ plane (the bar is placed along the $x$-axis), as compared to the total (thin+thick disc). We see that in models with both a thin and a thick disc the fraction of thin disc stars decreases as we go further away from the plane, while the fraction of the thick disc correspondingly increases.}
\label{fig:perc}
\end{figure}

We explore some of the morphological properties of bars and b/p bulges which form in setups with both a thin and thick disc. A full description of the morphological properties of bars and b/p bulges in thin+thick disc setups with different thick disc masses and scalelengths will be discussed in Fragkoudi et al. (in prep.); here we outline the most important aspects for this study.

In our simulations we can separate by construction which stars originate in the thin disc component, and which stars originate in the thick disc component, and we track these as the system evolves self-consistently. In what follows, we refer to the bar (or b/p) as seen exclusively in the thin disc population as the ``thin disc bar'' (or thin disc b/p), and that seen exclusively in the thick disc population as the ``thick disc bar'' (or thick disc b/p). 

In Figure \ref{fig:xysurf} we show the surface density of the thin (left) and thick disc (middle), and the total surface density for all the particles in the disc (right panel) at the end of the simulation after 9\,Gyr, once the bar and b/p have formed. We draw thick contour lines\footnote{Note that the thick contours are not for the same isophote level in the thin and thick disc} around isophotes of the bar in the left and middle panels, to draw the eye to some interesting features, namely that the thick disc bar is rounder than in the thin disc component. From ellipse fitting of the isodensity contours, we find that the thin disc bar has an axial ratio about half that of the thick disc at the end of the bar. This is not suprising, since previous studies have shown that bars forming in hotter components are weaker (e.g. \citealt{Combesetal1990, Athanassoula2003}). However it is interesting to note that even though the thin and thick discs are evolving concurrently in the same gravitational potential, the morphology of both the bar in the two components is different (see also \citealt{BekkiTsujimoto2011, BekkiTsujimoto2011b, DiMatteo2016, Debattistaetal2016, Athanassoulaetal2017}. See also \citealt{WozniakMichelDansac2009} on how a kinematically cold population born out of a gaseous component is mapped into a b/p bulge.) We show in Section \ref{sec:mapping} that this is due to the fact that the kinematically hotter component exchanges less angular momentum with the dark matter halo, as well as being due to the large radial excursions of the thick disc particles, which carry high angular momentum particles to the inner regions of the galaxy.
We also see that the morphology of the b/p is different in the two populations. The thin disc b/p has a much more prominent X-shape than the thick disc b/p, which looks rather more ``boxy'' (see also \citealt{BekkiTsujimoto2011, BekkiTsujimoto2011b, DiMatteo2016, Debattistaetal2016, Athanassoulaetal2017}). This again is due to the orbital structure of the two populations and the fact that the colder population is more efficiently trapped in the vertical resonances. 

The fact that the shapes of the bar and b/p are different in the thin and thick disc can also be seen more quantitatively by examining Figure \ref{fig:morph}, where we show the bar strength (left) and the b/p strength (right) for the thin and thick disc component, as well as for the total population. The bar strength is obtained using the now standard Fourier decomposition method (see e.g. \citealt{Athanassoulaetal2013}). We obtain the Fourier components by azimuthally decomposing the mass distribution as,

\begin{equation}
a_m(R) = \sum_{i=0}^{n_i} m_i \cos(m\theta_i), m = 0, 1, 2, ...
\label{eq:1}
\end{equation}

\begin{equation}
b_m(R) = \sum_{i=0}^{n_i} m_i \sin(m\theta_i), m = 0, 1, 2, ...
\label{eq:2}
\end{equation}

where $R$ is the radius, $n_i$ is the number of particles in an annulus around the radius $R$, $m_i$ is the mass of the particle and $\theta_i$ is the azimuthal angle. We perform this decomposition for the thin and thick discs separately, as well as for all the particles of the disc together.

A single value for bar strength is then obtained for each snapshot as,

\begin{equation}
A = \max (A_{2j}) = \max \Big( \frac{\sqrt{a^2_{2j}(R) + b^2_{2j}(R)}}{a_{0j}(R)}\Big)
\end{equation}

where $j$ stands for either the thin, thick or total mass distribution. 

The b/p strength is derived by taking the median of the absolute value of the distribution of particles in the vertical ($z$) direction, in a given radial bin of a snapshot seen edge-on, with the bar viewed side-on (similarly to \citealt{MartinezValpuestaAthanassoula2008}). To obtain the b/p strength we renormalise the median $\tilde{|z|}$ to that obtained at the start of the simulation $z_{0i}$, before the b/p formation and subsequent vertical heating. Then for each snapshot the maximum of these values over radius is taken as the boxy/peanut strength, i.e.,

\begin{equation}
C = \max \Big( \frac{\tilde{|z|}}{z_{0i}}\Big).
\label{eq:bps}
\end{equation}

In this way, the thin and thick disc will have the same b/p strength at the start of the simulation (this ensures that although the thick disc has a larger scaleheight at the beginning of the simulation, it will not have a ``stronger'' b/p). 

By examining Figure \ref{fig:morph}, we see that the strength of the thick disc bar is $\sim$50\% smaller than that of its thin disc counterpart, while the thick disc b/p strength is 20\% that of the thin disc b/p. This is due to the higher velocity dispersion in the thick disc, which leads to the stars being less tightly bound in the bar - b/p instability. On the other hand, the thin disc bar appears stronger because the particles in the thin disc are colder, and are thus on more circular orbits to begin with, and can therefore be more easily trapped in the bar instability.
We also see that the total bar and b/p strength, as well as the morphology of the bar and b/p of the total disc, are a combination of the two separate thin and thick disc components, i.e. they are intermediate between the two extremes, as seen in the third panel of Figure \ref{fig:xysurf} and the solid black lines in Figure \ref{fig:morph}. The degree to which the morphology will look like one or the other component will depend on how much mass is in each component, and on the relative scalelength of the two discs.

It is important to note that, due to the b/p being weaker in the thick disc component, the signature of the b/p will appear at larger heights above the plane for the thick disc stars than for the thin disc stars. By ``signature'' of the b/p, we refer to the dip in the density distribution of stars along the bar major axis (as is seen for example in the magnitude distribution of red giant clump stars in the Milky Way -- e.g. \citealt{Natafetal2010,McWilliamandZoccali2010,Nessetal2012,RojasArriagadaetal2014}). This can be seen in Figure \ref{fig:losdens}, where we take cuts in $z$ along the length of the bar (the bar is placed along the $x$-axis as in Figure \ref{fig:xysurf}) and plot histograms of the number of particles in the thin and thick disc. Very close to the plane, the signature of the b/p is not visible in either of the two components, and instead there is one central peak in the density distribution of stars (see Figure \ref{fig:los_dens_lowlow}). At slightly larger heights above the plane (Figure \ref{fig:los_dens_low}) the signature of the peanut begins to show up in the thin disc while it is not seen in the thick disc. However, as we go to larger heights above the plane (Figure \ref{fig:los_dens_high}) the peanut signature also appears in the thick disc, although the signature is weaker than in the thin disc. Additionally, the separation between the peaks of the peanut increases for larger heights above the plane (see for example also \citealt{Nessetal2012,WeggGerhard2013,Gonzalezetal2015,DiMatteo2016,Gomezetal2016, Debattistaetal2016}), as occurs also for the two peaks in the magnitude distribution of red clump stars along longitude $l$ =0 in the Milky Way (see for example \citealt{RojasArriagadaetal2014}). We see therefore that in order to be able to see the (weak) dip in the thick disc stars in these models, it is necessary to go to larger heights above the plane. 

In Figure \ref{fig:perc} we show the projection in $xz$ of the fraction of the thin disc surface density compared to the total. This shows in which areas of the disc above and below the plane the thin and thick disc component dominate the density distribution. The thin disc component dominates in the central regions close to the plane and is responsible for the strong X-shape of the b/p bulge. On the other hand the thick disc fraction dominates at regions further out above the plane. Thus we see that if we associate the thick disc with a metal poor component, and the thin disc with a metal rich component we would recover a vertical metallicity gradient in these models (see also \citealt{BekkiTsujimoto2011,BekkiTsujimoto2011b,DiMatteo2016,Debattistaetal2016}) as will also be shown in upcoming work using a model specifically adapted for comparison with the Milky Way.

\section{The los velocity of thin and thick discs in b/p's}
\label{sec:results}

\begin{figure}
\centering
\includegraphics[width=0.98\linewidth]{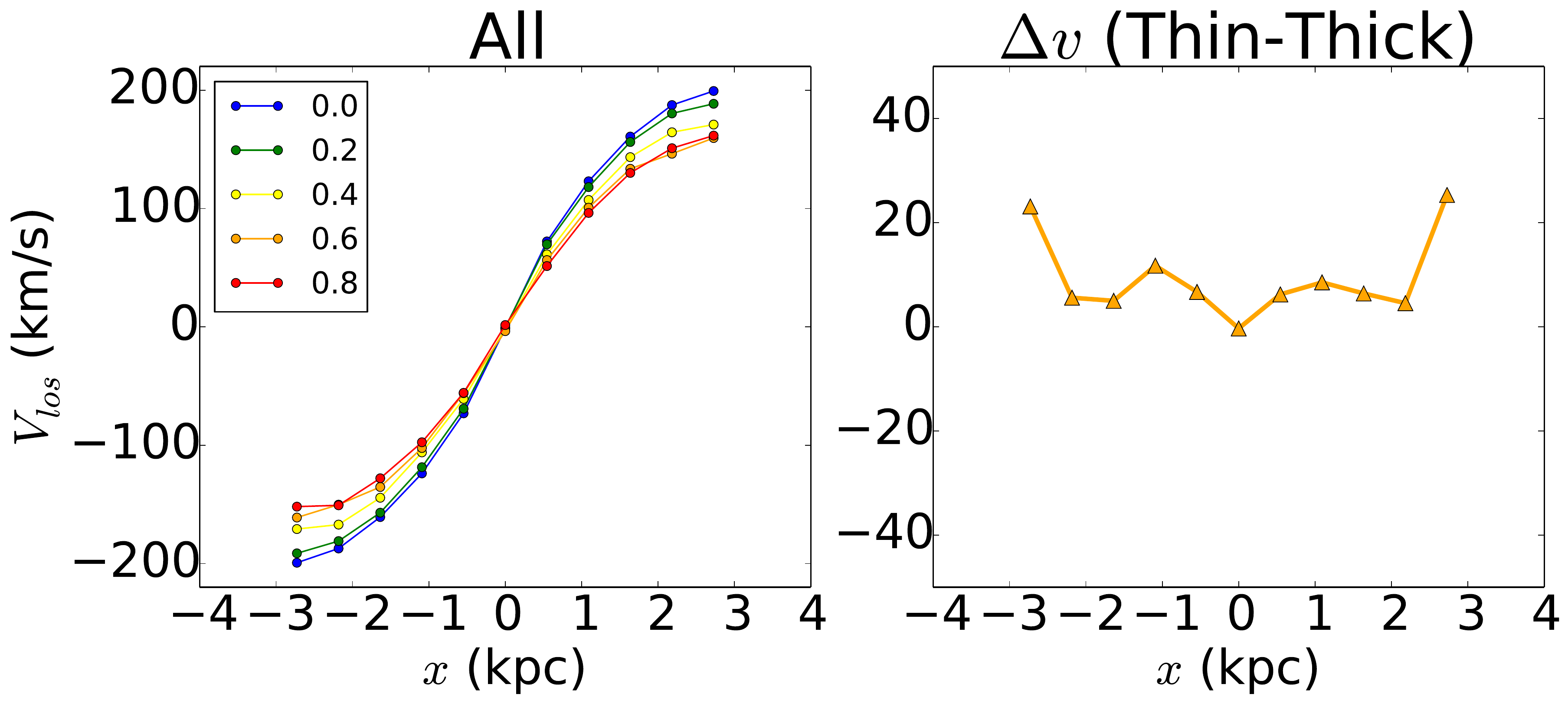}
\caption{\emph{Left:} The line of sight velocity of the disc in the central few kpc, (for all the particles) at the start of the simulation before the bar and b/p form, for different heights above the plane as indicated in the top left corner (values are in kpc). \emph{Right:} The difference between the absolute value of $v_{los}$ of the thin and thick disc components, $\Delta v$ (see Equation \ref{eq:3}). We see that the thin disc has higher los velocities than the thick disc at the start of the simulation before the bar - b/p forms.}
\label{fig:losv_001}
\end{figure}

\begin{figure}[h!]
\centering
\includegraphics[width=0.98\linewidth]{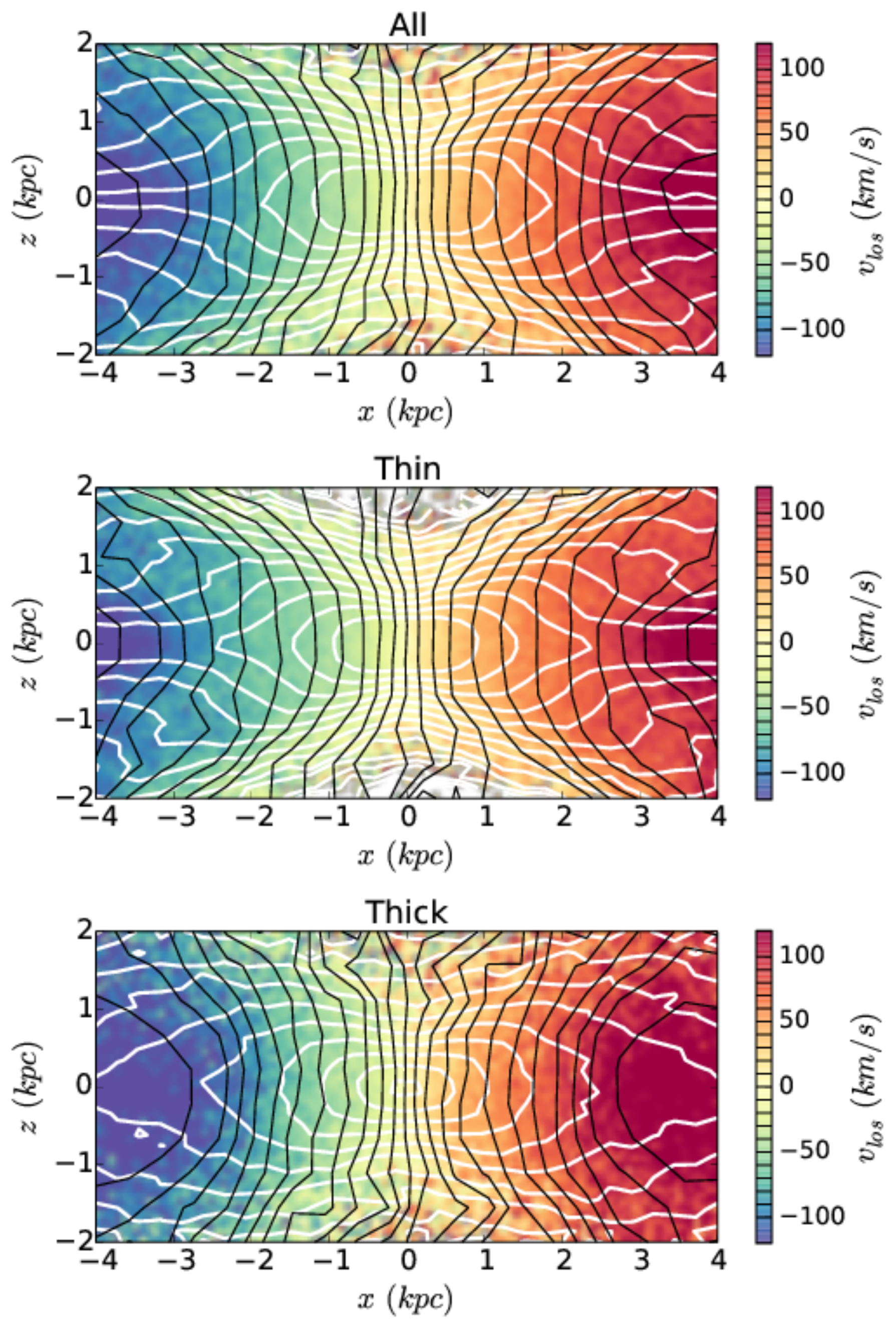}
\caption{2D velocity field of all the stars (top), of the thin disc stars (middle) and of the thick disc stars (bottom) in the fiducial simulation after 9\,Gyr of evolution, after the bar and b/p have formed. The white contours show the surface density of each component, while the black isovelocity contours are plotted at an interval of 10\,km/s.}
\label{fig:2dv}
\end{figure}

\begin{figure*}
\centering
\includegraphics[width=1.\linewidth]{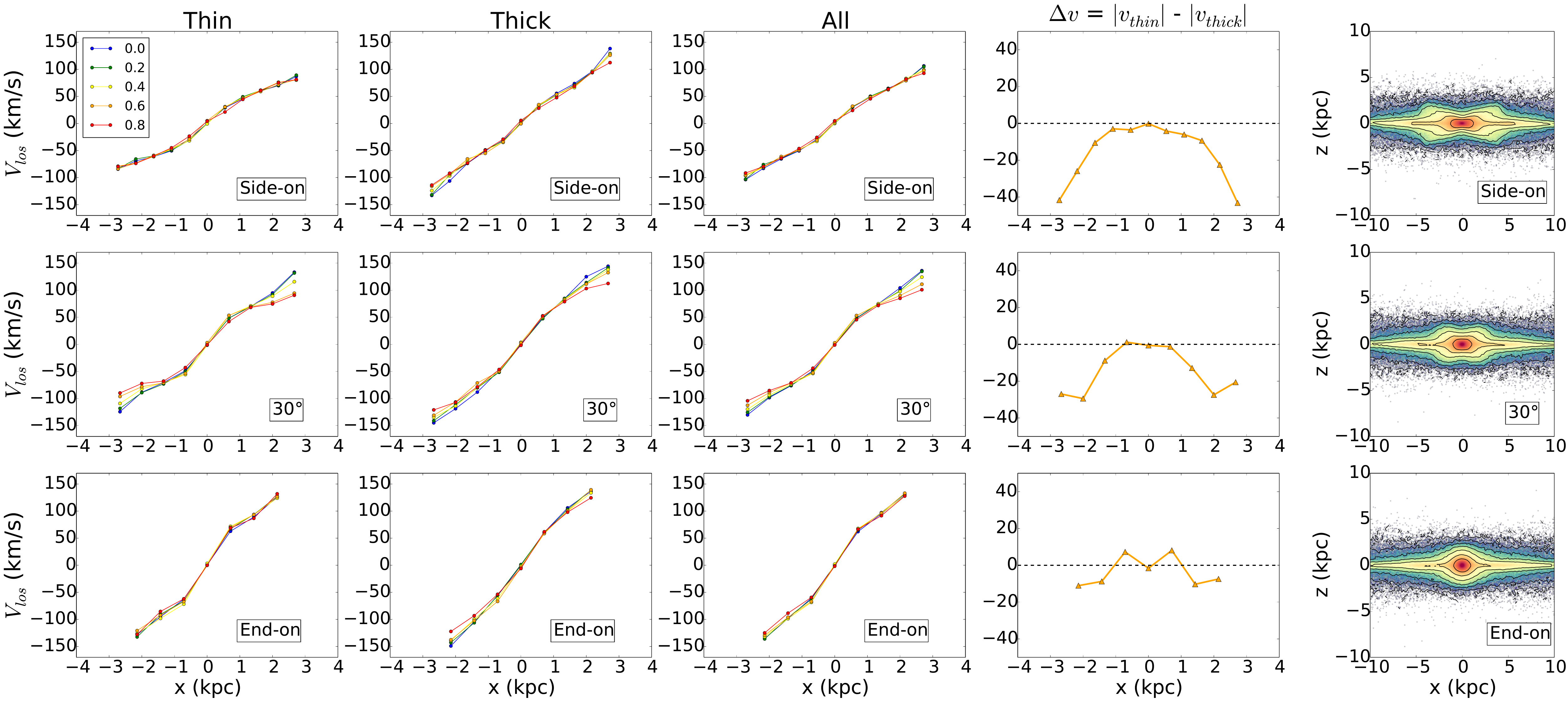}
\caption{The line-of-sight velocity for ``slits'' at different heights above the plane, as indicated in the top left inset, for three orientations of the bar, side-on (bar semi-major axis at 90 degrees to the los, top row),  30 degrees to the line of sight (similar to the Milky Way orientation - middle row) and end-on (bar at 0 degrees to the los, bottom row). The first column shows the los velocity of the thin disc stars, the second column the thick disc stars, and the third gives the los velocity for all the stars. The fourth column shows the average difference $\Delta v$ between the thin and thick disc los velocities (see Equation \ref{eq:3}). The fifth column shows the surface density of the simulation for the given orientation. We see that for side-on and Milky Way-like orientations the thick disc has higher los velocities than the thin disc.}
\label{fig:vlosm03}
\end{figure*}

In this section we explore the line-of-sight (los) velocity for this model of a disc galaxy with both a thin and a thick disc population. 

In the left panel of Figure \ref{fig:losv_001} we show the los velocity of all the particles in the disc for different heights above the plane, as indicated by the colours in the top left panel -- where $z$ is given in kpc -- at the start of the simulation, before the bar and the b/p form. We see that the galaxy does not exhibit cylindrical rotation in the central regions. In the right panel we show $\Delta v$,
\begin{equation}
\Delta v = |<v_{los,thin}>| - |<v_{los,thick}>| ,
\label{eq:3}
\end{equation}
where $v_{los}$ is the line of sight velocity of the thin and thick disc and $\Delta v$ is the difference between the absolute value of the average (for different heights above the plane) of these two line of sight velocities. 
By examining the right panel of Figure \ref{fig:losv_001} we see that, as expected, at the start of the simulation, the thin disc has a higher los velocity than the thick disc.
 
In Figure \ref{fig:2dv} we show a 2D map of the line of sight (los) velocity for all the particles (top), for the thin disc particles (middle) and for the thick disc particles (bottom) of the final snapshot of the fiducial simulation, after the bar and b/p are formed. The orientation of the bar is along the $x$-axis, i.e. side-on, and the white contours show the outline of the surface density of the corresponding component. As is typical of models containing bars and b/p bulges, there is overall cylindrical rotation in the model with the los velocity being independent of height above the plane (the reader is referred to \citealt{IannuzziAthanassoula2015} for a complete study of the 2D kinematics of the b/p bulge in models with a single disc). We also see that the thin and thick disc components also exhibit cylindrical rotation, until about $z\sim$1\,kpc above the plane of the galaxy.

We further explore the los velocity after the bar - b/p formation of each disc component separately in Figure \ref{fig:vlosm03}. We show the los velocity in the b/p bulge for different heights below the plane of the galaxy. We separate the particles into those originating in the thin and thick disc components (first and second columns respectively), and also plot the los velocity of all the particles together (third column). In the fourth column we show the difference in los velocities, $\Delta v$ (as given by Equation \ref{eq:3}), between the thin and thick disc. From top to bottom we show three different orientations of the bar: the side-on orientation where the bar is along the $x$-axis, an orientation in which the bar is at 30 degrees to the line of sight (similar to the view we have of the Milky Way) and end-on, i.e. where the bar is along the $y$-axis. The surface density in $xz$ of the stellar particles in each of these orientations is shown in the fifth column.

\begin{figure}
\centering
\includegraphics[width=0.9\linewidth]{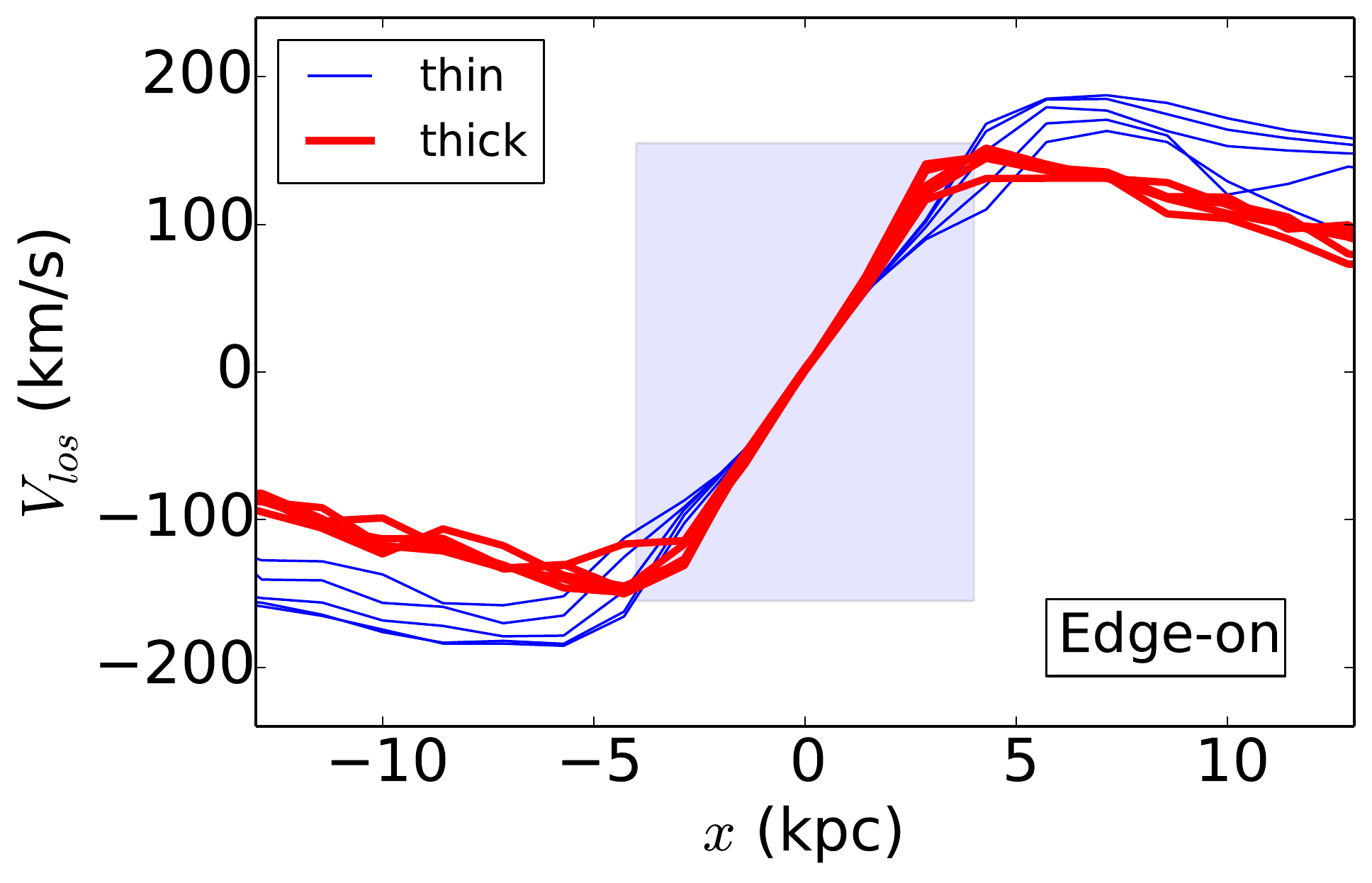}
\caption{The line of sight velocity of the thin (blue) and thick (red) disc stars at the end of the simulation, after the bar and b/p form. The shaded region indicates the zoomed-in region shown in Figure \ref{fig:vlosm03} and the lines correspond to different heights above the plane as in Figure \ref{fig:vlosm03}.}
\label{fig:losv_big}
\end{figure}

We see again that, apart from the cylindrical rotation in all the stars combined (third column of Figure \ref{fig:vlosm03}), also the thin and thick discs (1st and 2nd column of Figure \ref{fig:vlosm03}) exhibit cylindrical rotation \emph{separately}, as seen already in the 2D map of $v_{los}$ in Figure \ref{fig:2dv}. Additionally, we see that the $v_{los}$ depends on the orientation of the bar with respect to the line of sight. For end-on orientations, the los velocity has a steeper gradient than the side-on orientations. 
Interestingly, we see that the los velocity is not the same in the thin and thick disc components: we see that the thick disc stars have a \emph{higher} los velocity than thin disc stars in the outer regions of the b/p bulge, which is quantified by $\Delta v$ in the fourth column. This is contrary to what occurs at larger radii -- i.e. outside the b/p bulge region -- as can be seen by examining Figure \ref{fig:losv_big}, where we show a zoom-out of the $v_{los}$ of the thin and thick disc; the shaded box shows the area being examined in Figure \ref{fig:vlosm03}. The higher los velocity in the b/p region for thick disc stars is also contrary to what is seen at the start of the simulation before the bar forms (see right panel, Figure \ref{fig:losv_001}), where the thin disc has a higher los velocity. While the higher los velocity in the thick disc is most prominent for the side-on orientation of the bar (up to $\sim$40\,km/s), it can also be seen when the bar has an orientation of 30 degrees with respect to the line of sight towards the galactic centre -- i.e. similar to what is seen for the Milky Way -- ($\sim$20-30\,km/s) as well as in the end-on orientation ($\sim$ 10\,km/s).

\begin{figure}[h!]
\centering
\includegraphics[width=0.98\linewidth]{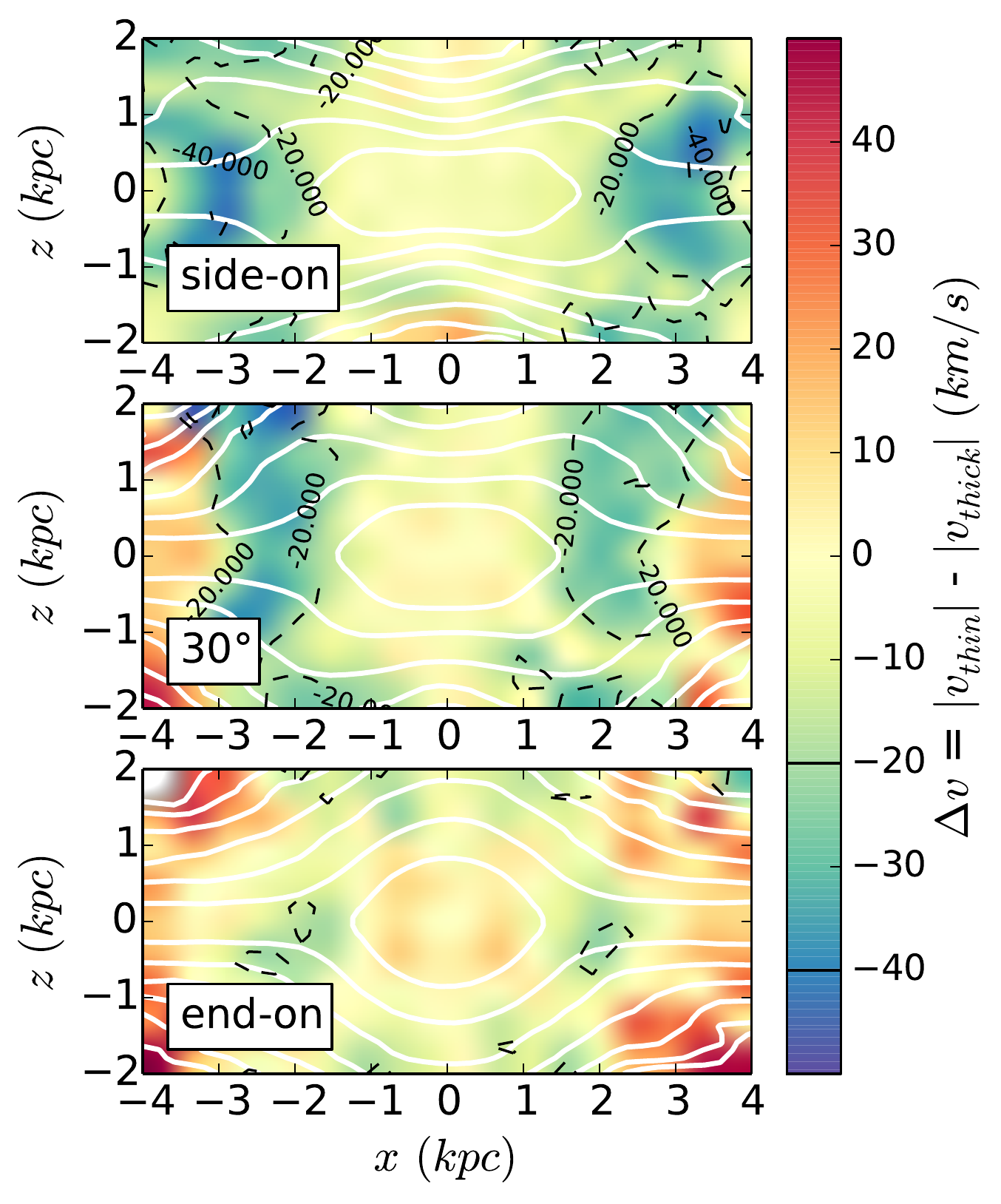}
\caption{The difference $\Delta v_{los}$ between the line of sight velocity of the thin and thick disc in 2D, for the three orientations shown in Figure \ref{fig:vlosm03}. Isodensity contours are shown in white.}
\label{fig:losv_2D_360}
\end{figure}

\begin{figure}
\centering
\includegraphics[width=0.98\linewidth]{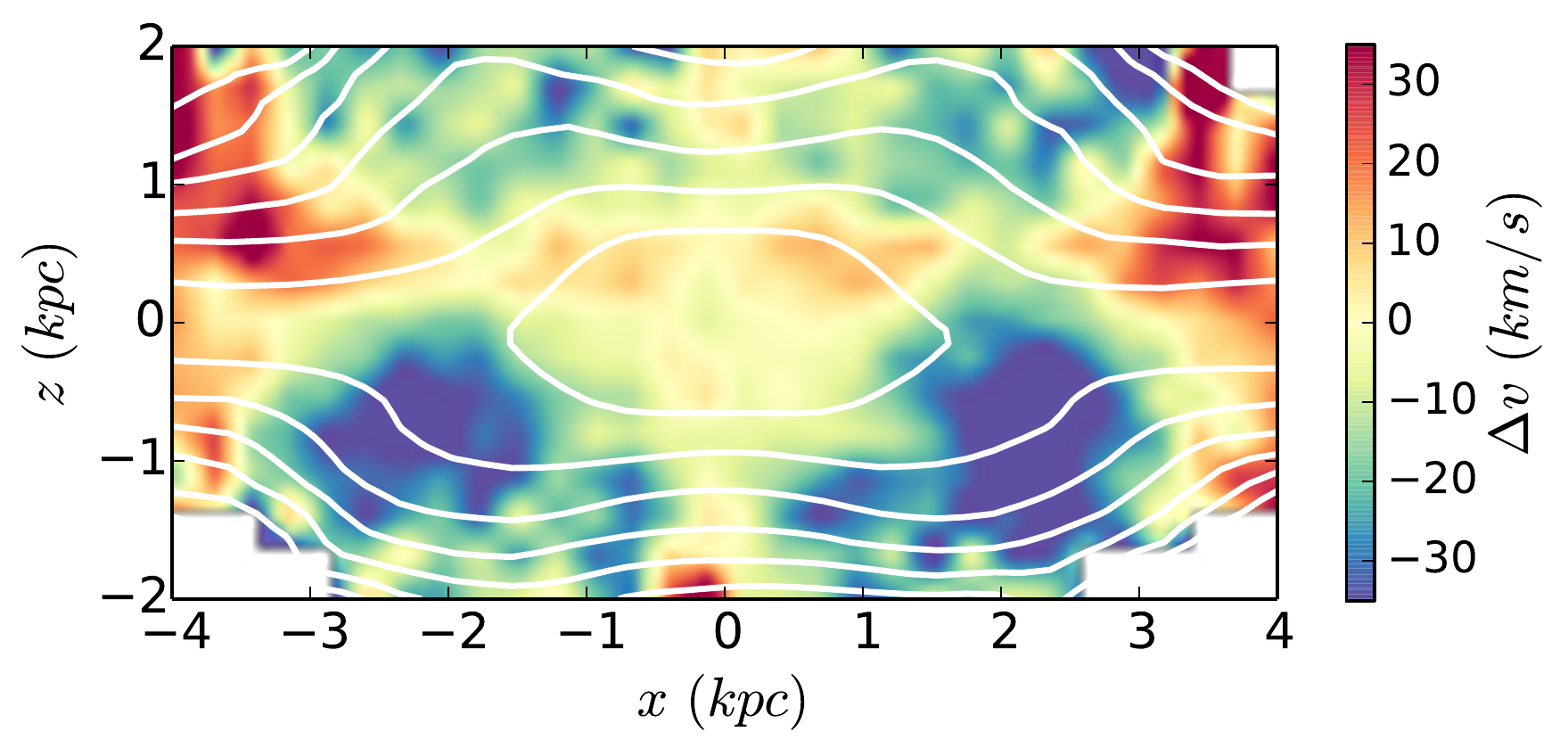}
\caption{The difference $\Delta v_{los}$ between the line of sight velocity of the thin and thick disc in 2D, for a snapshot in which the b/p in the model is not completely symmetric, as indicated by the white isodensity contours.}
\label{fig:losv_2D_250}
\end{figure}

We also show the difference between the thin and thick disc los velocity, $\Delta v_{los}$, in 2D in Figure \ref{fig:losv_2D_360} for the three different orientations mentioned above (from top to bottom: side-on, 30 degrees and end-on), where again the difference in $v_{los}$ in the two populations appears mostly at the edges of the b/p bulge. At the centre of the b/p, the thin disc can actually have a slightly higher $v_{los}$, especially for the end-on case. We also show $\Delta v$ between the thin and thick disc in Figure \ref{fig:losv_2D_250} for a snapshot in which the b/p is not completely symmetric yet (i.e. it is still going through the buckling phase). When the b/p is not completely symmetric there can be regions in the inner b/p where the thin disc has higher los velocity than the thick disc, while on the opposite side of the $z$=0 axis the thick disc will have a higher los velocity. 

Thus, in general in the \emph{outer} parts of the b/p the hot population will tend to have a higher los velocity, while in the inner regions of the b/p this trend can be reversed, depending also on the orientation of the bar with respect to the line of sight and on the symmetry of the b/p.

This is in fact a characteristic signature of a kinematically hot \emph{disc} component, since, while a dispersion dominated population -- such as a classical bulge -- can acquire some rotation, it cannot rotate \emph{faster} than the surrounding disc component (see for example \citealt{Fux1997,Sahaetal2012,SahaGerhard2013,DiMatteoetal2014,Sahaetal2016}).

\section{Mapping thin and thick discs in b/p's}
\label{sec:mapping}

\begin{figure}
\centering
\includegraphics[width=1.\linewidth]{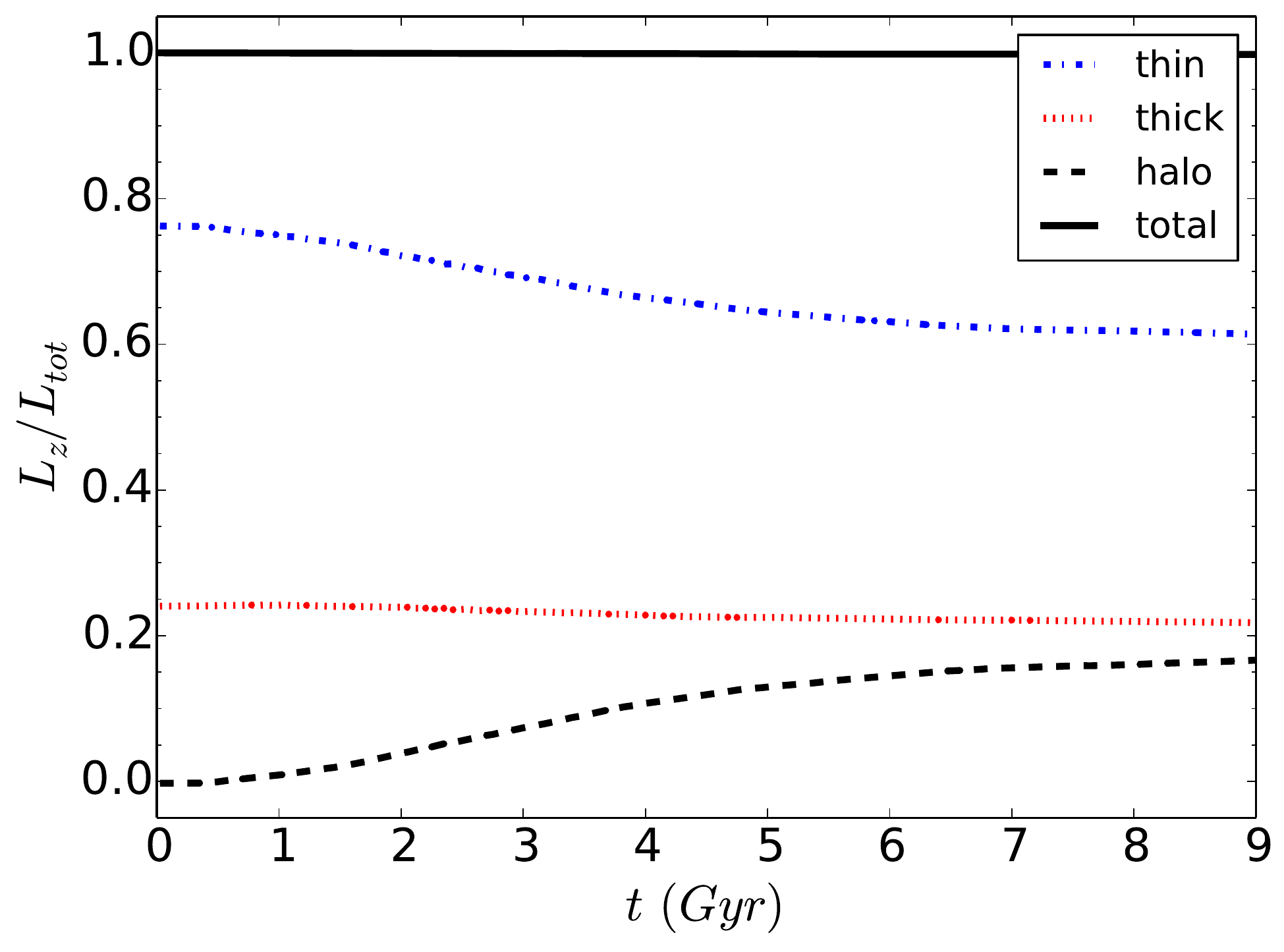}
\caption{The total angular momentum of the thin (dashed-dotted blue) and thick (dotted red) discs, of the dark matter halo (dashed black) and of the total angular momentum in all the components (solid black) as a function of time, normalised by the total angular momentum. The discs lose angular momentum which is transferred to the dark matter halo, while the thin disc loses a much larger percentage than the thick disc (see text).}
\label{fig:lzall}
\end{figure}

\begin{figure}
\centering
\includegraphics[width=0.98\linewidth]{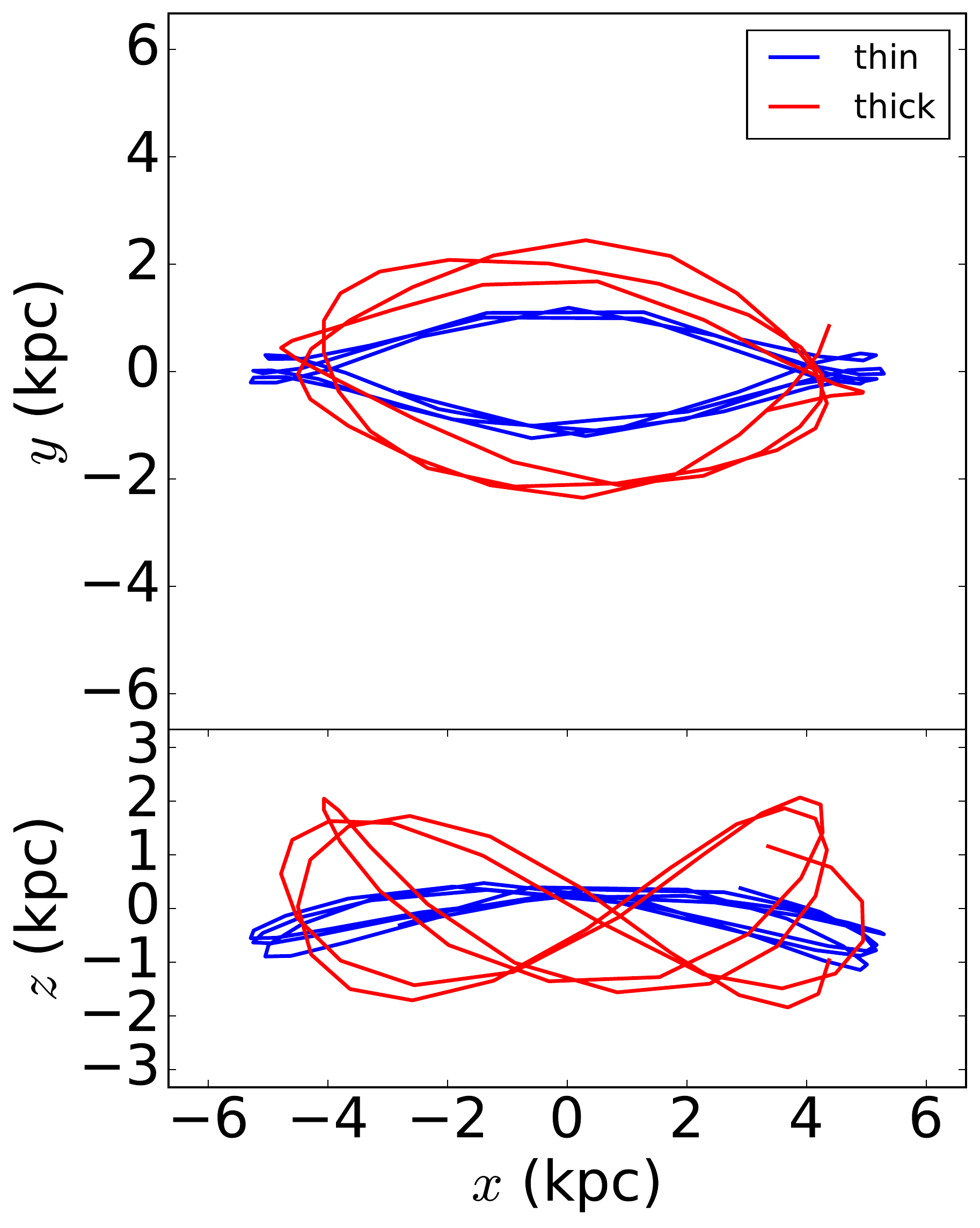}
\caption{Typical thin (blue) and thick (red) disc orbits in the bar region (see text for more details). The orbits look like the well-known $x_1$ orbits in the face-on projection (the bar is rotated to be along the $x$-axis). For similar radii, thick disc orbits tend to be rounder than thin disc orbits. }
\label{fig:orbs}
\end{figure}

\begin{figure*}
\centering
\includegraphics[width=0.98\linewidth]{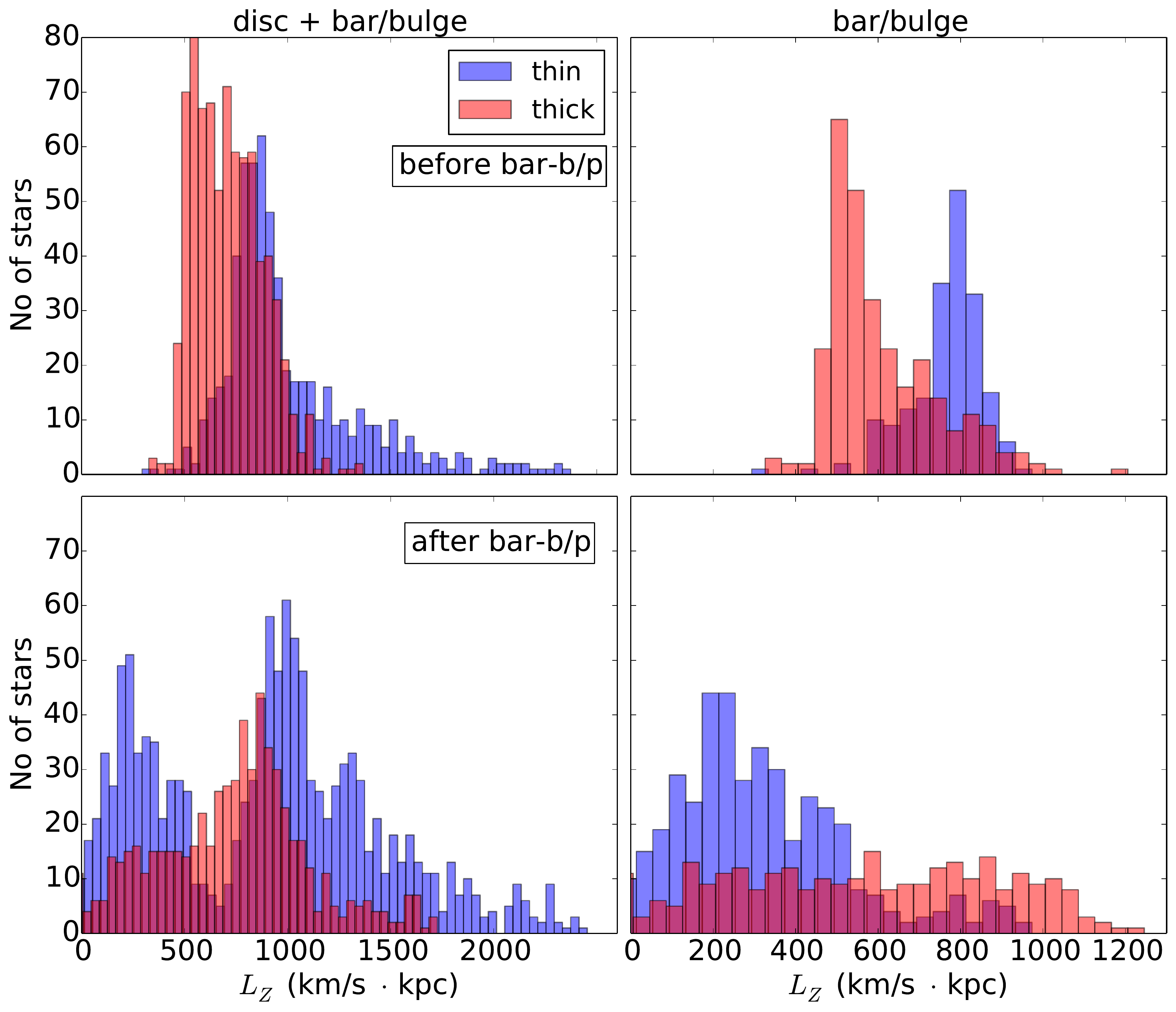}
\caption{The angular momentum distribution of the thin (blue) and thick (red) disc particles. \emph{Left:} The particles are chosen from a region including the disc and bar, i.e. $x$ = 3.2 - 3.5\,kpc, $z$ = 0.2 - 0.5\,kpc and $y$= -8 - 8\,kpc (unshaded region of Figure \ref{fig:xysurf}). In these plots the foreground and background disc contamination are explicitly included. \emph{Right:} The angular momentum distribution of the thin (blue) and thick (red) disc particles, in the bar-b/p region, i.e. $x$ = 3.2 - 3.5\,kpc, $z$ = 0.2 - 0.5\,kpc and $y$ = -1 - 1\,kpc (shaded region of Figure \ref{fig:xysurf}). The top row shows the angular momenta for the first snapshot when the disc is axisymmetric, i.e. before the bar and b/p bulge form, while the bottom row shows the angular momenta after bar and b/p formation. Focusing on the particles in the bar/bulge region after the bar and b/p form (bottom right panel) we see that the thick disc particles have a tail of higher angular momentum than the thin disc stars.}
\label{fig:lz}
\end{figure*}

\begin{figure*}
\centering
\subfigure[Thin disc stars]{%
	\includegraphics[width=0.45\linewidth]{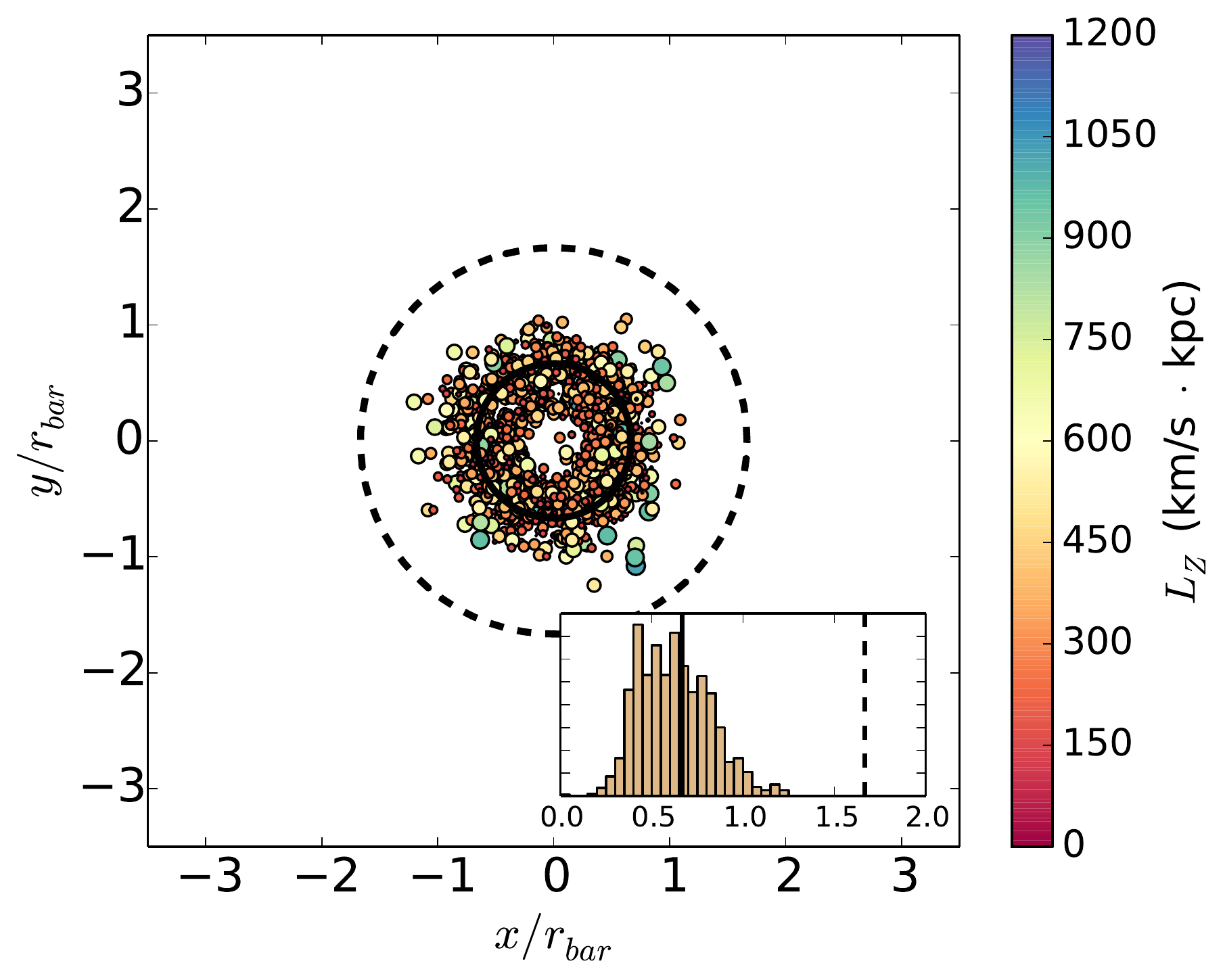}
	\label{fig:birththin}}
\quad
\subfigure[Thick disc stars]{%
	\includegraphics[width=0.45\linewidth]{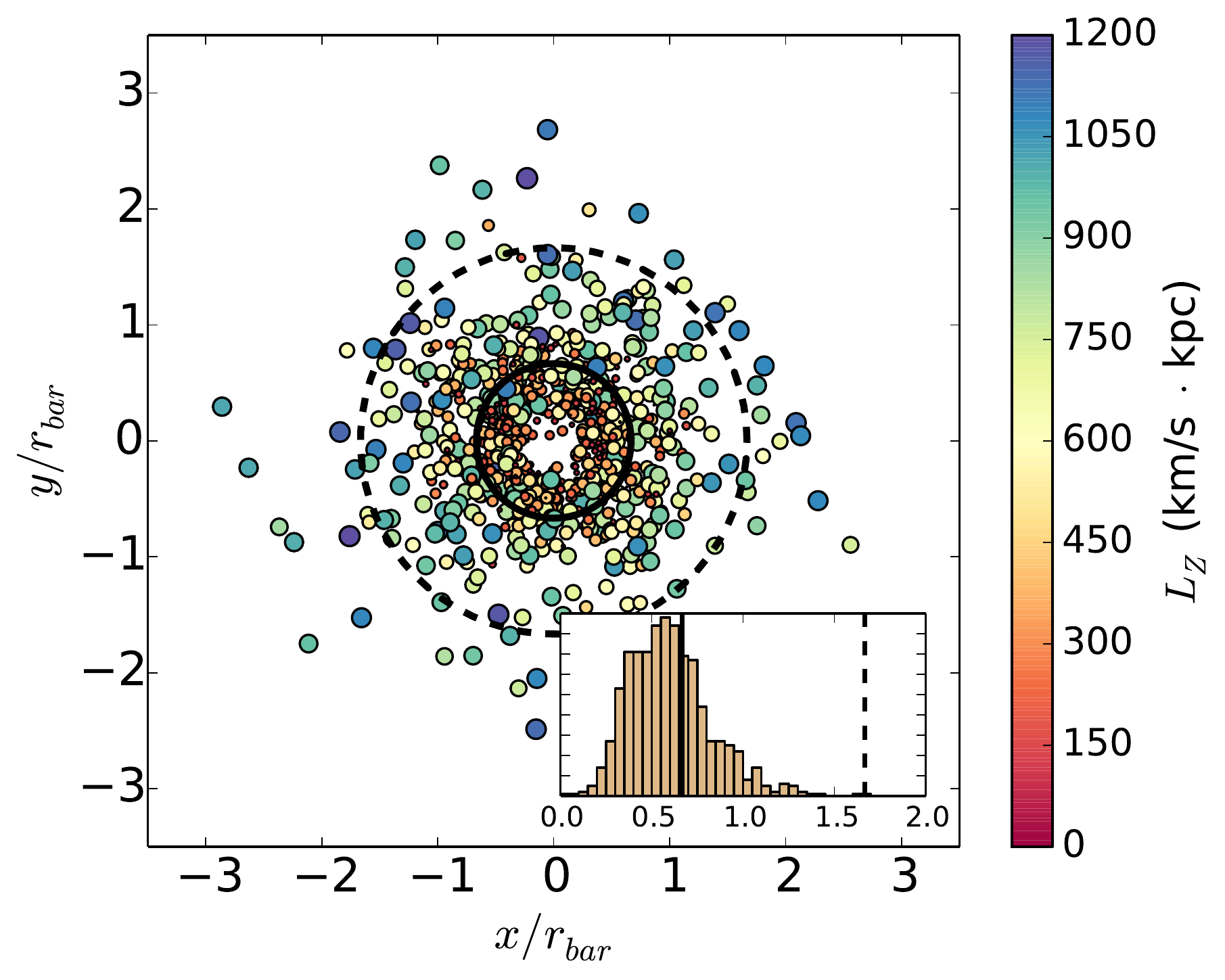}
	\label{fig:birththick}}
\quad
\caption{The instantaneous position of particles selected near the edge of the bar (shaded regions of Figure \ref{fig:xysurf}) at the start of the simulation for the thin (left) and thick (right) disc. The $x$ and $y$ axis are normalised to the bar length $r_{bar}$. The solid black circle gives the average radius of these stars at the end of the simulation and the dashed black circle delineates the OLR. The inset shows a histogram of the guiding radii of these stars at the start of the simulation. The guiding radii of the thin disc stars are quite similar to the instantaneous radius of the particles, while the guiding radii of the thick disc stars are limited to a smaller range than the instantaneous positions, since these stars have larger radial excursions. The size and colour of the points indicate the angular momentum of these particles at the end of the simulation. We see that the stars born at larger radii are also the ones that have larger angular momenta at the end of the simulation.}
\label{fig:birth}
\end{figure*}

In this section we explore why thick disc stars have a higher los velocity in the b/p bulge region, and how this is related to the orbits (and therefore also morphology), angular momenta and birth radii of stars found in the bar - b/p region, and relate these to the radial velocity dispersion of the thin and thick disc components.

\subsection{Angular momentum exchange}

The kinematics of the thin and thick disc stars in the bar - b/p bulge region can be understood in terms of the amount of angular momentum that the stars of the thin and thick disc lose via the bar instability. As has been extensively studied in the literature (e.g. \citealt{Athanassoula2003,CeverinoKlypin2007}), the disc transfers angular momentum to the halo mainly via the corotation, inner and outer Lindblad resonances which are induced by the bar potential. In Figure \ref{fig:lzall} we plot the angular momenta (normalised to the total angular momentum $L_{tot}$ of the system) of the various components in the simulation as a function of time. The thin disc, which is kinematically colder than the thick disc, transfers angular momentum to the halo, which subsequently increases its angular momentum. The thick disc's angular momentum on the other hand slightly increases during the first 1.5\,Gyrs of evolution (by about 1\%), after which it also proceeds to lose angular momentum via the bar instability which is transferred to the dark matter halo, although at a much slower rate than the thin disc. This initial increase in the angular momentum of the thick disc could be due to a transfer of angular momentum from the thin disc. Overall, the dark matter halo's angular momentum increases by about 16\% $L_{tot}$, while the thin and thick discs' angular momenta decrease by about 14\% and 2\%$L_{tot}$ respectively. The decrease in angular momentum is thus more significant for thin disc than for thick disc stars.
The amount of angular momentum which is transferred from the discs to the dark matter halo is related to the radial velocity dispersions of the discs. Stars from a colder population with a lower radial velocity dispersion will get trapped more easily by the bar instability and thus the transport of angular momentum from the disc to the halo will be more efficient. 

The fact that the higher $v_{los}$ of thick disc stars is most evident for side-on orientations of the bar, is indicative of the fact that the bar orbits are responsible for this behaviour. Indeed, as we can see by examining the morphology of the thin and thick disc bar in Figure \ref{fig:xysurf}, focusing on the thick isodensity contours, the thick disc bar is rounder than the thin disc bar. This can also be understood in terms of the models' orbital structure, as can be seen in Figure \ref{fig:orbs}, where we plot typical orbits found in the bar region, which originate from the thin and thick disc. These orbits are taken by selecting the stars in the shaded region of Figure \ref{fig:xysurf} and tracing the orbits over a short period of time, while maintaining the bar along the $x$-axis. We then examined these orbits by eye and found a common trend, i.e., that thick disc bar orbits in general tend to be rounder than thin disc orbits with the same semi-major axis, as the example shown in Figure \ref{fig:orbs}. We will show a full orbital analysis of thin and thick disc stars in bar and b/p bulges in upcoming work.

To further explore this, we examine the specific angular momentum of stars in the outer parts of the b/p bulge (the selected stars are indicated by the shaded and unshaded coloured regions in Figure \ref{fig:xysurf}). We show, in Figure \ref{fig:lz}, the angular momenta of stars selected from this region at the start of the simulation when the disc is axisymmetric (top row) and at the end of the simulation once the bar and b/p form (bottom row). In the left column we plot the angular momenta of \emph{all} the stars along the line of sight (i.e. we include the foreground and background disc contamination) in the region of $x$ = 3.2 - 3.5\,kpc and $z$ = 0.2 - 0.5\, kpc (indicated by the unshaded boxed region of Figure \ref{fig:xysurf}), while in the right column we only select stars in the bar region (as indicated by the shaded regions in Figure \ref{fig:xysurf}). 

We see some interesting features in the angular momentum distribution: first, as expected, at the start of the simulation the thin disc stars have higher angular momenta than the thick disc stars (top left plot). Also at the start of the simulation, at a radius that will eventually be the end of the bar - b/p region, the thin disc stars also have higher angular momenta than the thick disc stars, with a clear peak of the angular momentum of the thin disc stars at higher values of $L_z$ (top right plot). 
At the end of the simulation, once the bar and b/p have formed, we see a bimodality in the angular momenta of the thin disc stars in the bottom left panel of Figure \ref{fig:lz}: there is a peak at low angular momenta (around 250 km/s$\times$kpc) and a peak at high angular momenta (around 1000 km/s$\times$kpc), while the angular momenta of the thick disc stars have spread out and the peak is shifted to slightly higher angular momenta compared to the start of the simulation. The low angular momentum peak of stars in the thin disc is due to the fact that the thin disc loses a significant of angular momentum from the inner regions, and ends up with stars on  \emph{elongated bar orbits}; The peak at higher angular momenta for the thin disc stars is due to the stars in the disc on largely circular orbits, outside the bar. This is confirmed by examining the angular momenta of thin disc stars in the bar region (bottom right plot), in which we see that once we exclude the disc from the selection, we maintain only the low angular momentum stars in the bar. We also see (in the bottom right plot of Figure \ref{fig:lz}) that in the bar region, the thick disc stars have a tail of high angular momentum. These stars have higher angular momenta partly because thick disc stars do not lose as much angular momentum via the bar instability as thin disc stars do (see Figure \ref{fig:lzall}) and partly because, as we discuss below, thick disc stars from large radii are trapped in the bar instability. 

\subsection{Birth radii}

In addition to the fact that thin disc stars lose more angular momentum than thick disc stars, we find that the tail of high angular momentum stars in the thick disc is also due to stars which originate further out in the disc, but due to their large radial excursions can be found in the inner regions of galaxies, in the bar-b/p region. 

To explore this, in Figure \ref{fig:birth} we plot the ``birth radii'' i.e. the positions of the particles at the start of the simulation, of the selected particles from the shaded regions of Figure \ref{fig:xysurf}. The solid and dashed black lines indicate the average radius of the particles and the Outer Lindblad Resonance (OLR) at the end of the simulation. The colour and size of the points in the plot are a function of the angular momentum at the end of the simulation, i.e. after the bar - b/p formation. It's clear that particles maintain a ``memory'' of their original angular momentum since stars which are born at larger radii also tend to have the largest angular momentum at the end of the simulation (see also \citealt{MartinezValpuestaGerhard2013,DiMatteoetal2014}). We see in Figure \ref{fig:birththin} that the thin disc particles originate from a region around the average radius, with some migration from both inside and outside this radius. When examining the birth radii of thick disc particles in Figure \ref{fig:birththick} which end up in the b/p region, we see that there are a number of stars with birth radii much larger than those of the thin disc. This is due to the fact that thick disc stars have larger radial excursions, and thus can end up in the inner parts of the disc; thus these stars have high angular momenta at the end of the simulation. These large radial excursions are due to the higher radial velocity dispersion in the outer parts of the thick disc. 

This migration is mostly due to blurring, rather than churning, due to the fact that stars in the thick disc are on elongated orbits; this is consistent with the fact that we expect churning to be less efficient for hot populations \citep{VeraCiroetal2014}. This is corroborated by examining the insets of Figure \ref{fig:birth}, where we show the guiding radius of the particles at the start of the simulation, which are calculated as in \cite{Minchevetal2014}, where the guiding radius is given by,

\begin{equation}
r_{g} = \frac{L}{v_{c}},
\end{equation}

where $L$=$rv_{\phi}$ is the angular momentum of the particle and $v_c$ is the circular velocity at the corresponding radius.

We see that the guiding radii of the thick disc particles are in general smaller than the instantaneous birth radii and that all these stars have guiding radii inside the OLR (see also \citealt{Halleetal2015}), while stars can have instantaneous birth radii outside the OLR. Therefore we see that thick disc stars enter the bar - b/p region due to blurring, i.e. due to the fact that the particles are on elongated orbits.


\subsection{Dependance on thick disc velocity dispersion}
\label{sec:depsr}
\begin{figure}
\centering
\includegraphics[width=0.85\linewidth]{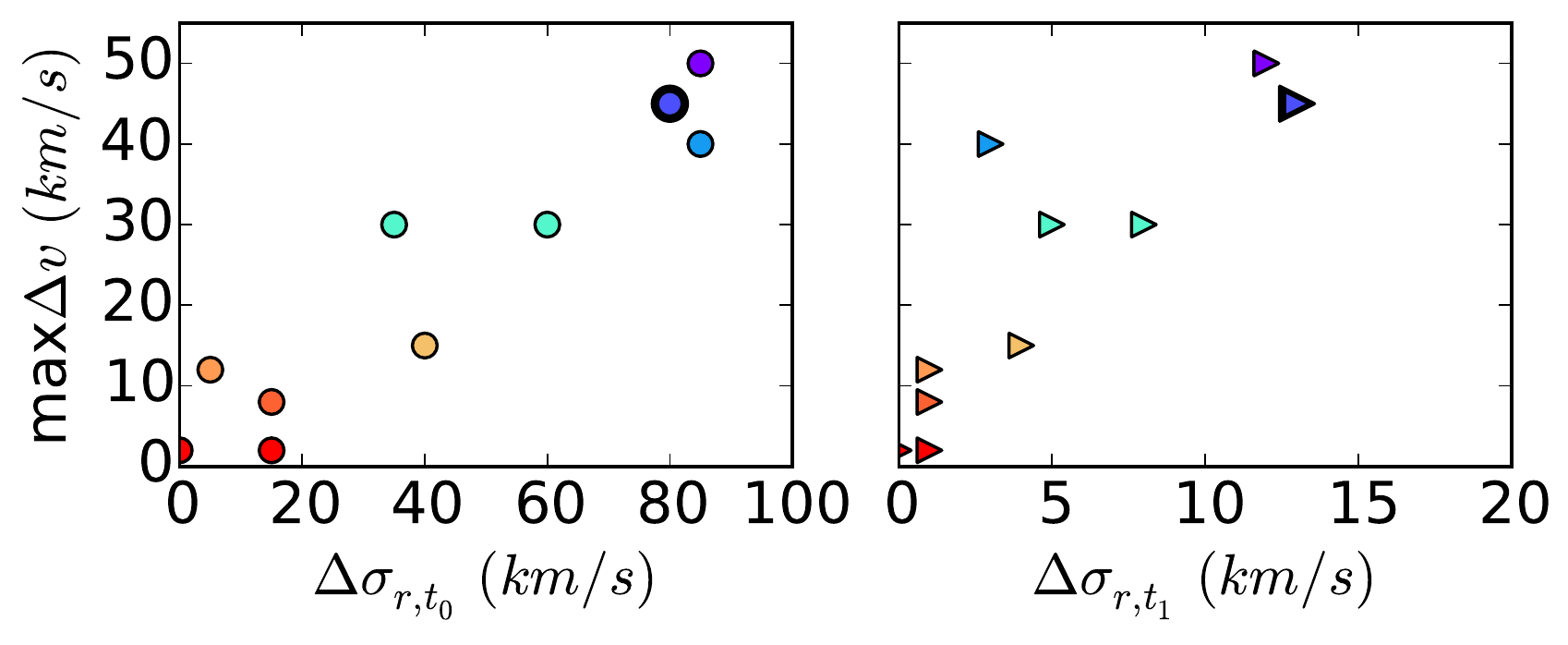}
\caption{\emph{Left:} Dependance of the difference in los velocity between thin and thick disc stars in the b/p as a function of the difference in radial velocity dispersion at $r$=$r_{b/p}$, at the start of the simulation. The colour coding is as a function of the $y$-axis and is added to help guide the eye in Figure \ref{fig:sigmar_tests}. The model analysed in this paper is indicated by a black outline. \emph{Right:} Same as left panel but for the final snapshot of the simulation.}
\label{fig:sigmardep}
\end{figure}

The radial velocity dispersion profiles of discs at the time of bar and b/p formation influences two factors which will determine their los velocity in the bar - b/p region: i) the angular momentum transfer from the discs to the halo through the bar instability and ii) the radii from which stars can reach the bar - b/p region. By increasing the radial velocity dispersion in the thick disc there is less angular momentum transfer from the disc to the halo, and stars can reach the bar - b/p region from larger radii; these factors will lead to higher los velocities of thick disc stars compared to the thin disc stars in the bar - b/p regions. 

In terms of what we expect the radial velocity dispersion profile of the thick disc to be, there is little observational evidence, both at low as well as at high redshifts, when bars and boxy/peanut bulges first formed. There have been some recent studies from the CALIFA survey which show that the velocity dispersion for late type galaxies seems to increase in their outer parts (see Figure 10 of \citealt{FalconBarrosoetal2016}). However this velocity dispersion is the los dispersion and does not correspond only to the radial velocity dispersion, $\sigma_r$, and it also corresponds to the velocity dispersion at low redshifts -- not the velocity dispersion at the time that bars and b/p bulges might have formed in these galaxies.
From the theoretical side, cosmological zoom-in simulations, such as the one studied in \cite{Birdetal2013}, show that the oldest stellar populations, which are also the hottest, can have quite high radial velocity dispersions in the outer parts, with a dip in velocity dispersion at intermediate radii and then rising again towards the centre (see their Figure 4). Therefore, the shape of the thick disc radial velocity dispersion profile produced in our fiducial initial conditions seems to not be uncommon at least in simulated galaxies. 

Nonetheless, since we do not know what the radial velocity dispersion profile of galaxies will be at redshifts z$\sim$1-2 -- when bars are expected to start forming in discs (e.g. \citealt{Shethetal2008,Simmonsetal2014,Gadottietal2015}) -- we explore the effect of changing this profile on the los velocity of thin and thick disc stars in the b/p bulge. We can thus determine what kind of trends we can expect if the radial velocity dispersion of the thin and thick disc were different to begin with (see also \citealt{Debattistaetal2016} for a discussion on the effects of $\sigma_r$ on the morphology of the b/p bulge). To this aim, we carried out a number of test simulations, in which we imposed various radial velocity dispersion profiles for the thick disc (see Figure \ref{fig:sigmar_tests} in the Appendix).

We find that whether or not the thick disc stars have a higher los velocity compared to the thin disc stars will depend (at least in part) on the shape of the initial velocity dispersion profile (i.e. before bar formation). Specifically, it depends on the difference in velocity dispersion between the thin and the thick disc, in the regions from which particles eventually get trapped in the bar and b/p bulge.
If the radial velocity dispersion of the thin and thick disc are similar, then the morphology of the bar will also be similar in the two discs and we would not expect to see a higher los velocity in the thick disc stars. Indeed there needs to be quite a large difference between the initial radial velocity dispersions of these discs in order to observe the difference in $v_{los}$, as can be seen in the left panel of Figure \ref{fig:sigmardep}, where we plot the maximum difference between the average los velocity of the thin and thick disc, $\Delta v$, as a function of the difference between the velocity dispersion in the thin and thick discs $\Delta \sigma$ at the start of the simulation (at a radius corresponding to the edge of the b/p at the end of the simulations) for a number of test simulations. We see that there is a general trend, in which the larger the difference between the radial velocity dispersions in the thin and thick disc at the start of the simulation, the larger the difference in the los velocities once the bar and b/p forms. We also plot the difference in velocity dispersion of the thin and thick disc at the radius where we measure $\Delta v$ at the end of the simulation, for reference, and we show this in the right panel of Figure \ref{fig:sigmardep}. We see that the trend remains but that the difference in velocity dispersion between the cold and hot population is reduced.

This of course indicates that the properties of the bar and b/p in thin and thick discs will be highly dependent on the properties of the discs at the time of bar formation/disc settling. Thus if thick discs are formed with higher velocity dispersions than thin discs -- which is what observations \citep{Wisnioskietal2015} and simulations \citep{Birdetal2013,Grandetal2016} of galaxies at higher redshifts point to -- the morphology and kinematics of the thick disc bar will be considerably different to that of the thin disc bar. We suggest, very tentatively, that it could be possible to use the relation found in Figure \ref{fig:sigmardep}, to place constraints on the difference between the velocity dispersions of the thin and thick disc at the time the bar was formed.

\section{Comparison with observations}
\label{sec:discussion}

In this section we compare the findings of this study to some of the available observations of the morphology and kinematics of the Milky Way bulge, even though the model presented in this paper is not fitted to represent the Milky Way. Do we see these kinds of trends, i.e. a higher (or equal) los velocity in kinematically hotter components than in the kinematically colder ones in the b/p bulge of the Milky Way? How does the morphology of the b/p bulge change with height above the plane and for different stellar populations?

\subsection{Morphology}

Using data from the ARGOS survey \citep{Freemanetal2013}, which explored the metallicity and kinematics of stars in the inner 3.5\,kpc of the Milky Way bulge, Ness et al. separate the stars in their sample into different components according to their metallicity (Ness et al. 2013a,b)\nocite{Nessetal2013a,Nessetal2013b}. They dub the main components they find in the Milky Way bulge as A, B and C, where A is the most metal rich component with [Fe/H] > 0, component B with 0 > [Fe/H] > -0.5 and component C, -0.5 > [Fe/H] > -1. They also find a small fraction of stars in the bulge with lower metallicities, dubbed components D and E, with [Fe/H] < -1, which make up just a few percent of the stars in the bulge. 

The fraction of each component changes with height above the plane, which leads to a vertical metallicity gradient \citep{Nessetal2013a}; close to the plane the metal rich component A has a larger contribution to the total density, while further away from the plane, at $b$ = -10 degrees, its contribution decreases and that of the metal poor component C, increases. Component B, seems to be more or less the same at all heights above the plane. Although in our models we only have two components -- a thin and thick disc which could be thought of as a metal rich and metal poor component respectively -- we see that having a metal poor thick disc and a metal rich thin disc would give qualitatively similar results, i.e. a higher fraction of the metal rich component close to the plane and a higher fraction of the metal poor component further out from the plane (see Figure \ref{fig:perc}). 

In the Gaia-ESO Survey (GES) \citep{Gilmoreetal2012}, the stars in the bulge are separated into two components, metal poor and metal rich, where the metal-poor component is also kinematically hotter than the metal-rich component \citep{RojasArriagadaetal2014}. It can be seen in Figure 10 of \cite{RojasArriagadaetal2014}, that for the field closest to the plane (p1m4 at $b$ = -4), there is no dip in the magnitude distribution of the red clump giants in their sample. However, for fields further away from the plane (e.g. at p0m6, $b$ = -6) there is a clear dip in the magnitude distribution of the red clump stars of the metal rich component, which is however not visible in the metal-poor component. For field m1m10, which is at $b$ = -10, and therefore the furthest away from the plane, the dip in the metal rich component becomes wider, while there seems to be a hint of a dip in the metal poor component as well (which in \citealt{RojasArriagadaetal2014} is interpreted as being due to a volume effect). This dip at higher latitudes for the metal-poor component is very similar to what we see in our Figure \ref{fig:losdens}, in which we take cuts parallel to the $z$ plane and explore the density along the los for the thin and thick disc component. That is, we find that the signature of the b/p bulge in the kinematically hotter (thick) disc component appears weaker and further away from the plane than for the kinematically cold (thin) disc population. In \cite{RojasArriagadaetal2014} there is also a faint dip in the field p7m9, which is at 7 degrees of longitude and which is likely not due to the b/p bulge. We note that while the bimodality at p7m9 may be spurious, we suspect that the one at m1m10 is not, since this signature is seen both in GES and in ARGOS. In the ARGOS data (Figure 20 of \citealt{Nessetal2013a}) we see a bimodality in population B at $l$ = 0, $b$ = -10. Since the metal poor population of GES includes part of what would correspond to population B of ARGOS, and since the ARGOS $l$ = 0, $b$ = -10 field is very close to the GES m1m10 field, it is not unreasonable to suggest that the bimodality in the metal-poor population in m1m10 in GES is not spurious. However, as the model used in this work is not specifically adapted to fit the MW bulge we do not explore this further and we refer to future work in which we will use a precise model for the MW to explore in detail the signature of the b/p bulge in terms of $l$,$b$.

\subsection{Kinematics}

In the ARGOS survey, component A is the kinematically coldest component, and the components become hotter for B and C, while maintaining cylindrical rotation \citep{Nessetal2013b}. Components D and E on the other hand do not have cylindrical rotation. This seems to point to the fact that populations A, B and C likely have a disc-like origin. We also see that component B, which is kinematically hotter than component A, has a \emph{higher} los velocity than component A in the outer regions of the Bulge, similarly to what we find in this study, which indicates that component B is likely due to a kinematically hotter \emph{disc}. This is similar to the interpretation given by \cite{Nessetal2013b} for component B (although not due to the los velocity), who find that the chemical properties of population B are similar to those of the ``early thin disc'' and which in \cite{DiMatteoetal2014} and \cite{DiMatteo2016} was interpreted as the ``young thick disc''. 
Regarding component C from \cite{Nessetal2013b}: although it has a higher velocity dispersion, it does not appear to have a higher los velocity. This could be due to another origin for component C, i.e. other than a thick disc, while it could also be due to some contamination from halo stars which do not have a disc-like origin (see for example \citealt{Debattistaetal2016}) and therefore do not have much rotation to begin with. 

We also see similar behaviour in the GES data \citep{RojasArriagadaetal2014}, in which, as mentioned above, the metal rich and metal poor components are also kinematically cold and hot respectively. We see a hint (although within the errors) of a higher los velocity in their metal-poor population as compared to the metal rich population in the bulge (see their Figure 11). This behaviour is to be expected, should this population be due to the thick disc population which can participate in the bar instability.

On the other hand, kinematic data for the bulge region from the APOGEE survey \citep{Nessetal2016} (in which the data is separated as in the ARGOS survey, into components A, B, C, etc.) does not seem to recover the same behaviour in terms of the los velocity as in the aforementioned surveys, with component B and C from APOGEE both rotating \emph{slower} than component A. It remains to be understood why this discrepancy exists between the different surveys and whether this just an issue of low number statistics in some fields.


\section{Summary \& Conclusions}
\label{sec:summary}

We examine the morphology and kinematics of the bar-b/p bulge region in N-body simulations of isolated disc galaxies with both a (kinematically cold) thin, and a (kinematically hot) thick disc, which evolve secularly over 9\,Gyr. 
By construction, we can separate the particles in our models into those originating from the thin and thick disc, and can therefore trace how the two discs are mapped into the bar - b/p. Due to the different kinematic properties of these two discs at the time the bar and b/p form, their stars have different morphological and kinematic properties when mapped into the bar and b/p.

In terms of morphology, we find that:
\begin{itemize}
\item The bar that forms out of stars from the kinematically cold thin disc is stronger ($\sim$50\%) and has an axial ratio which is almost half that of the thick disc bar (see Figures \ref{fig:xysurf} and \ref{fig:morph}).
\item The b/p bulge forming out of the thin disc is stronger and has a more pronounced X-shape than the thick disc b/p bulge (see Figures \ref{fig:xysurf} and \ref{fig:morph}). 
\item The signature of the b/p, i.e. the dip in the density distribution along the bar major axis, is weaker for thick disc stars and appears at larger heights above the plane compared to the thin disc stars (see Figure \ref{fig:losdens}).
\item The distance between the peaks of the b/p increases for larger heights above the plane (see Figure \ref{fig:losdens}).
\end{itemize}
In terms of kinematics, we see that:
\begin{itemize}
\item The model has cylindrical rotation, both in the total stellar population (i.e. thin + thick disc stars) and also in each component separately (see Figures \ref{fig:2dv} and \ref{fig:vlosm03}).
\item Contrary to what is seen in the disc region outside the b/p (Figure \ref{fig:losv_big}), the thick disc stars have \emph{higher} los velocities than those of the thin disc in the outer regions of the b/p, when the bar is viewed side-on or with an orientation similar to the one in the Milky Way (40\% and 20\% higher $v_{los}$ respectively; see Figure \ref{fig:vlosm03}). This feature is a characteristic signature of a kinematically hot \emph{disc} participating in the bar - b/p instability.
\item This feature is due to the orbital structure of the thin and thick disc bars, and the fact that stars from the thin disc get trapped on more elongated orbits in the bar region (see Figure \ref{fig:orbs}).
\item The orbital structure in the thin and thick disc bars is due to the angular momentum transfer from the discs to the halo. A considerable amount of angular momentum is transferred from the thin disc to the halo, while the thick disc does not lose as much angular momentum as the thin disc, due to the fact that it is a kinematically hotter population (see Figure \ref{fig:lzall}). 
\item Additionally, thick disc stars originating in the outer parts of the disc with high angular momentum, can be found in the bar region due to large radial excursions i.e. due to blurring (see Figures \ref{fig:lz} and \ref{fig:birth}). 
\item We find a correlation between the difference in the initial radial velocity dispersion of the discs ($\Delta\sigma_r$) and the difference in the los velocity of the two discs in the b/p bulge region at the end of the simulation ($\Delta v_{los}$; see Figure \ref{fig:sigmardep}). 
\end{itemize}

The aforementioned all point to the fact that the morphological and kinematic properties of bars and b/p bulges in galaxies with different stellar populations will be highly dependent on the kinematic properties of these populations -- and therefore also on their formation mechanisms.

\begin{acknowledgements}
This work has been supported by the ANR (Agence Nationale de la Recherche) through the MOD4Gaia project (ANR-15-CE31-0007, P.I.: P. Di Matteo). FF is supported by a postdoctoral grant from the Centre National d'Etudes Spatiales (CNES). This work was granted access to the HPC resources of CINES under the allocation 2016-040507 made by GENCI. 
\end{acknowledgements}

\bibliographystyle{aa}
\bibliography{References}%

\begin{appendix} 
\section{Radial velocity dispersions of test simulations}
\label{sec:appendixB}
\begin{figure*}[h!]
\centering
\subfigure[Thin]{%
	\includegraphics[width=0.45\linewidth]{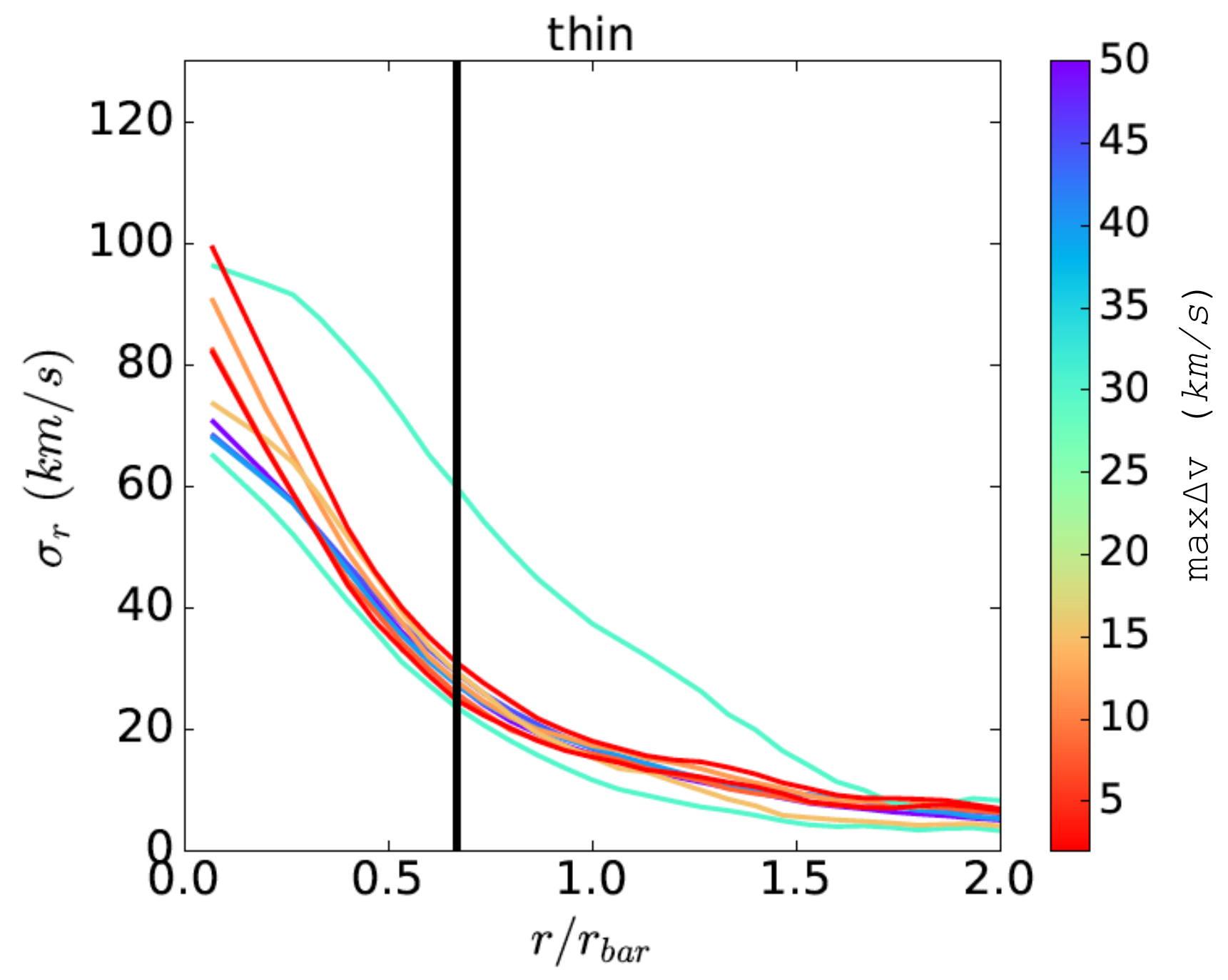}
	\label{fig:alldispthin}}
\quad
\subfigure[Thick]{%
	\includegraphics[width=0.45\linewidth]{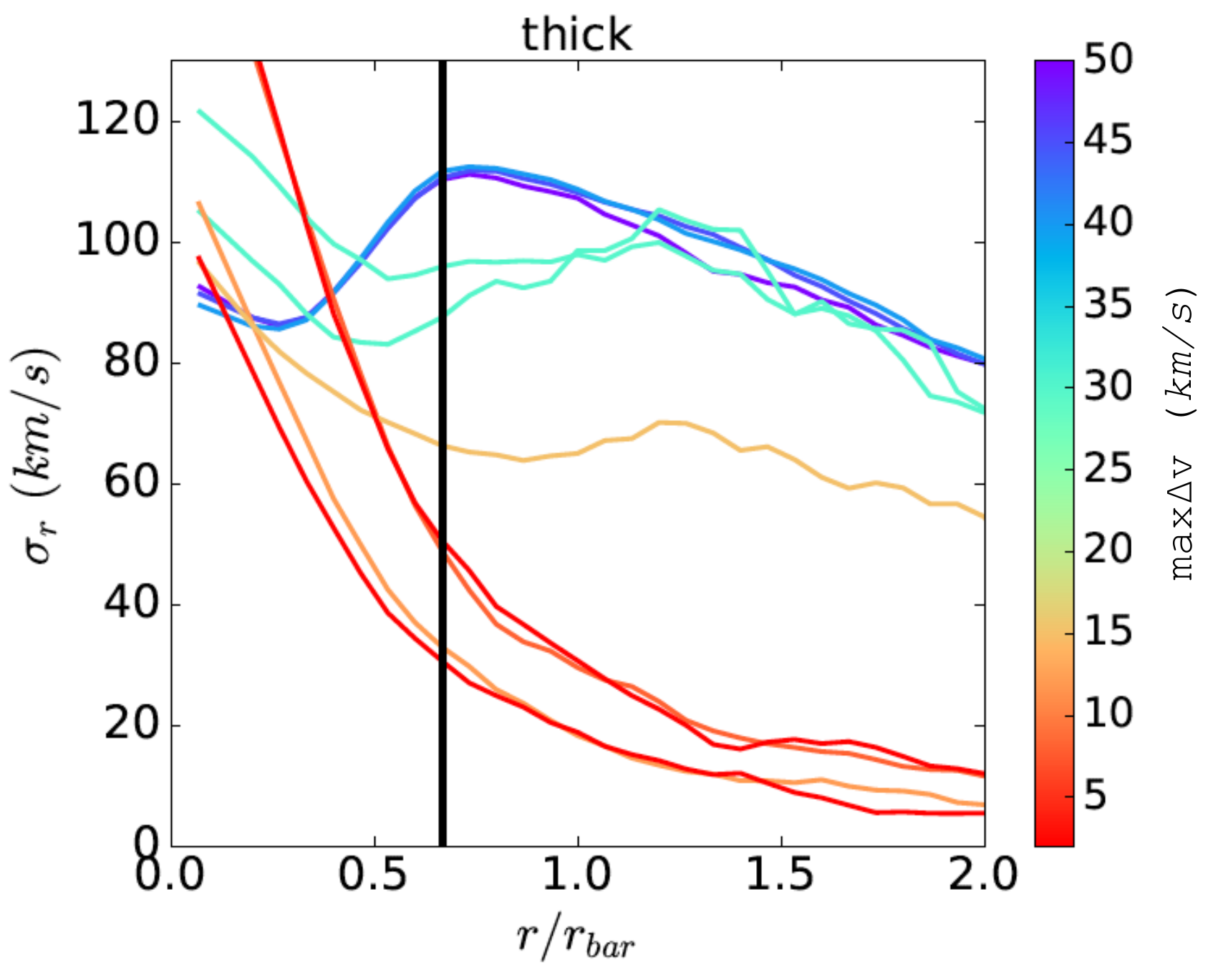}
	\label{fig:alldispthick}}
\quad
\caption{The initial radial velocity dispersion $\sigma_r$ of the thin (left) and thick disc (right) for a number of test simulations, which were used to construct Figure \ref{fig:sigmardep}. The colour coding corresponds to the colourbar in Figure \ref{fig:sigmardep}, i.e. according to the maximum $\Delta v$ of the model. The vertical black line indicates the end of the b/p bulge. }
\label{fig:sigmar_tests}
\end{figure*}
For the analysis discussed in Section \ref{sec:depsr} we used a number of test simulations with a variety of shapes for the radial velocity dispersion profile of the thick disc, in order to examine how the results discussed in the paper depend on this profile. We show in Figure \ref{fig:sigmar_tests} the radial velocity dispersion profiles of the test models used to construct Figure \ref{fig:sigmardep}, where the colour-coding of Figure \ref{fig:sigmar_tests} corresponds to that of Figure \ref{fig:sigmardep}. The vertical black line indicates the end of the b/p bulge radius, which is where the maximum difference between the los velocities of the thin and thick disc (max$\Delta_{v}$) can be seen in Figure \ref{fig:vlosm03}. This is where the difference between the radial velocity dispersion of the thin and thick disc were taken for Figure \ref{fig:sigmardep}.
\FloatBarrier

\end{appendix}

\end{document}